\documentclass[sigconf]{acmart}

\AtBeginDocument{%
  }

\copyrightyear{2022}
\acmYear{2022}
\setcopyright{acmlicensed}
\acmConference[KDD '22]{Proceedings of the 28th ACM SIGKDD Conference on Knowledge Discovery and Data Mining}{August 14--18, 2022}{Washington, DC, USA}
\acmBooktitle{Proceedings of the 28th ACM SIGKDD Conference on Knowledge Discovery and Data Mining (KDD '22), August 14--18, 2022, Washington, DC, USA}
\acmPrice{15.00}
\acmDOI{10.1145/3534678.3539125}
\acmISBN{978-1-4503-9385-0/22/08}

\usepackage{graphicx}
\usepackage{amsmath}
\usepackage{multirow}
\usepackage{booktabs}
\usepackage{array}
\usepackage{subfigure}
\usepackage{amssymb}
\usepackage{algorithm}
\usepackage{verbatim}
\usepackage{bm}     
\usepackage{enumitem}       

\usepackage{algpseudocode}

\begin{document}
\fancyhead{}

\title{Contrastive Cross-domain Recommendation in Matching}


\author{Ruobing Xie}
\authornote{Both authors have equal contributions. Ruobing Xie is the corresponding author.}
\affiliation{\institution{WeChat, Tencent}
\city{Beijing}
\country{China}}
\email{ruobingxie@tencent.com}

\author{Qi Liu}
\authornotemark[1]
\affiliation{\institution{WeChat, Tencent}
\city{Beijing}
\country{China}}
\email{addisliu@tencent.com}

\author{Liangdong Wang}
\affiliation{\institution{WeChat, Tencent}
\city{Beijing}
\country{China}}
\email{ldwang@tencent.com}

\author{Shukai Liu}
\affiliation{\institution{WeChat, Tencent}
\city{Beijing}
\country{China}}
\email{shukailiu@tencent.com}

\author{Bo Zhang}
\affiliation{\institution{WeChat, Tencent}
\city{Beijing}
\country{China}}
\email{nevinzhang@tencent.com}

\author{Leyu Lin}
\affiliation{\institution{WeChat, Tencent}
\city{Beijing}
\country{China}}
\email{goshawklin@tencent.com}


\begin{abstract}
Cross-domain recommendation (CDR) aims to provide better recommendation results in the target domain with the help of the source domain, which is widely used and explored in real-world systems. However, CDR in the matching (i.e., candidate generation) module struggles with the data sparsity and popularity bias issues in both representation learning and knowledge transfer. In this work, we propose a novel Contrastive Cross-Domain Recommendation (CCDR) framework for CDR in matching. Specifically, we build a huge diversified preference network to capture multiple information reflecting user diverse interests, and design an intra-domain contrastive learning (intra-CL) and three inter-domain contrastive learning (inter-CL) tasks for better representation learning and knowledge transfer. The intra-CL enables more effective and balanced training inside the target domain via a graph augmentation, while the inter-CL builds different types of cross-domain interactions from user, taxonomy, and neighbor aspects. In experiments, CCDR achieves significant improvements on both offline and online evaluations in a real-world system. Currently, we have deployed our CCDR on WeChat Top Stories, affecting plenty of users. The source code is in \url{https://github.com/lqfarmer/CCDR}.
\end{abstract}




\maketitle

\section{Introduction}
\label{sec.introduction}

Personalized recommendation aims to provide attractive items for users according to their profiles and historical behaviors, which has been widely implemented in various fields of our lives. Real-world large-scale recommendation systems usually adopt the classical two-stage architecture containing ranking and matching. The \emph{matching} module \cite{xu2018deep,xie2020internal,xie2021improving} (i.e., candidate generation \cite{covington2016deep}) focuses more on the efficiency and diversity, which first retrieves a small subset of (usually hundreds of) item candidates from the million-level large corpora. Next, the \emph{ranking} module gives the specific ranks of items for the final display.

\begin{figure}[!hbtp]
\centering
\includegraphics[width=0.96\columnwidth]{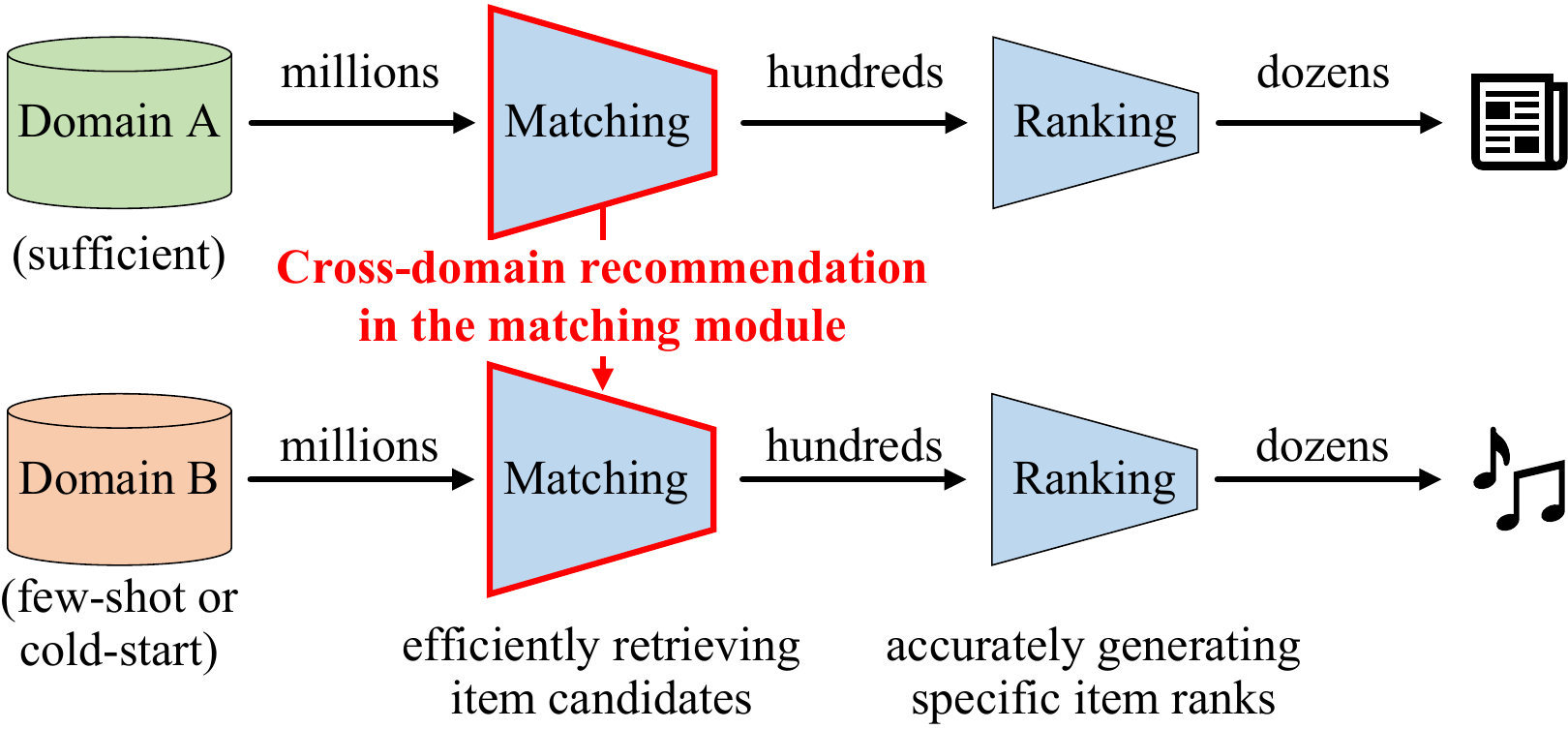}
\caption{An example of CDR in matching.}
\label{fig:example}
\end{figure}

With the increase of recommendation scale and the expansion of recommendation scenarios, real-world recommendations usually need to bring in additional data sources (i.e., domains) as supplements to improve their content coverage and diversity. These cold-start items of new data sources only have very few user behaviors at their warm-up stage. Hence, it is difficult to recommend these cold-start items appropriately.
\emph{Cross-domain recommendation (CDR)}, which aims to make full use of the informative knowledge from the source domain to help the target domain's recommendation \cite{zhu2021transfer}, is proposed to solve this issue. EMCDR \cite{man2017cross} is a classical CDR method, which focuses on building user mapping functions via aligned user representations in the source and target domains. CoNet \cite{hu2018conet} proposes another approach that jointly models feature interactions in two domains via a cross connection unit.
However, existing CDR methods often heavily rely on aligned users for cross-domain mapping (e.g., EMCDR), ignoring other rich information in recommendation such as taxonomy. It will harm the knowledge transfer between different domains, especially in cold-start scenarios. Moreover, lots of CDR methods are designed for ranking that consider complicated cross-domain user-item interactions (e.g., CoNet), which cannot be directly adopted in matching due to the online efficiency. CDR in the matching module should consider not only recommendation accuracy, but also diversity and efficiency.

In this work, we aim to improve the matching module's performance on new (few-shot or strict cold-start) domains via the CDR manner. Fig. \ref{fig:example} shows an illustration of this task. Precisely, CDR in matching mainly has the following three challenges:

(1) \emph{How to address the data sparsity and popularity bias issues of CDR in matching?} Real-world recommendation usually suffers from serious data sparsity issues when modeling the interactions between million-level users and items. Moreover, these sparse interactions are even highly skewed to popular items with high exposure owing to the Matthew effect \cite{paterek2007improving}, which makes hot items become hotter. These two issues inevitably harm the representation learning of cold-start and long-tail items, whose damages will even be multiplied in matching where all items should be considered.

(2) \emph{How to conduct more effective knowledge transfer for the (cold-start) target domain with few user behaviors?} As stated above, conventional CDR methods strongly depend on aligned users and their behaviors. The performance of CDR in matching will be greatly reduced, if most users and items have few interactions and models cannot learn reliable representations in cold-start domains. Moreover, other heterogeneous information (e.g., taxonomy) should also be fully considered in CDR to bridge different domains. We should build more effective and robust cross-domain knowledge transfer paths to well learn both popular and long-tail objects.

(3) \emph{How to balance the practical demands of accuracy, diversity and efficiency of CDR in matching?} Online efficiency requirements need to be strictly followed. Moreover, matching is more responsible for the diversity than ranking, for it determines the inputs of ranking. A good CDR matching model should comprehensively transfer user diverse preferences via multiple paths to the target domain.

To address these issues, we propose a novel \textbf{Contrastive Cross-Domain Recommendation (CCDR) } to transfer user preferences in matching.
Specifically, we build two global diversified preference networks for two domains, containing six types of objects to enhance diversity and cross-domain connections. We conduct a GNN aggregator with a neighbor-similarity based loss on heterogeneous interactions to capture user diverse interests.
To strengthen the cross-domain knowledge transfer, we design the \textbf{intra-domain contrastive learning (intra-CL)} and \textbf{inter-domain contrastive learning (inter-CL)} in CCDR. The intra-CL conducts an additional self-supervised learning with sub-graph based data augmentations to learn more reliable representations for matching in the target domain. In contrast, the inter-CL designs three contrastive learning tasks focusing on the cross-domain mapping between aligned users, taxonomies, and their heterogeneous neighbors. The mutual information maximization with different types of objects multiplies the effectiveness of cross-domain knowledge transfer.
Finally, all additional CL losses are combined with the original CDR losses under a multi-task learning (MTL) framework.
We conduct a cross-domain multi-channel matching to further improve the diversity in online.
CCDR has the following three advantages:
(1) The intra-CL brings in self-supervised learning for long-tail users and items, which can alleviate the data sparsity and popularity bias issues in matching.
(2) The inter-CL introduces new CL-based cross-domain interactions, which enables more effective and robust knowledge transfer for CDR in matching.
(3) The diversified preference network, multiple CL tasks, and the cross-domain multi-channel matching cooperate well to capture user diverse preferences, which meets the requirements of diversity and efficiency in online system.

In experiments, we compare CCDR with competitive baselines on real-world recommendation domains, and achieve significant improvements on both offline and online evaluations. Moreover, we also conduct some ablation tests and parameter analyses to better understand our model. The contributions are concluded as:
\begin{itemize}
\item We propose a novel contrastive cross-domain recommendation for CDR in matching. To the best of our knowledge, we are the first to conduct contrastive learning to improve representation learning and knowledge transfer in CDR.
\item We propose the intra-CL task with a sub-graph based data augmentation to learn better node representations inside the single domain, which can alleviate the data sparsity issue in CDR of matching.
\item We also creatively design three inter-CL tasks via aligned users, taxonomies, and their neighbors in our diversified preference networks, which enable more effective and diverse knowledge transfer paths across different domains.
\item We achieve significant improvements on both offline and online evaluations. CCDR is simple, effective, and easy to deploy. It has been deployed on a real-world recommendation system for more than $6$ months, affecting plenty of users.
\end{itemize}

\section{Related Works}

\textbf{Matching in Recommendation.}
The matching (i.e., candidate generation) module aims to retrieve a small subset of (usually hundreds of) items from large corpora efficiently \cite{xie2020internal}.
Recently, embedding-based retrieval \cite{huang2020embedding,cen2020controllable} is also widely used in practical systems with the help of fast retrieval servers \cite{johnson2019billion}. Due to the need for efficiency, embedding-based methods usually adopt a two-tower architecture, which conducts two different towers for building user and item representations separately. Different feature interaction modeling methods such as FM \cite{rendle2010factorization}, YoutubeDNN \cite{covington2016deep}, AutoInt \cite{song2019autoint}, ICAN \cite{xie2020internal}, and AFT \cite{hao2021adversarial} could be used in these towers.
In contrast, tree-based matching models \cite{zhu2018learning} give another way to address the matching problem with structured item trees. Graph-based matching models \cite{xie2021improving} are also proposed to learn user/item representations.
However, few works focus on the CDR in matching, which often exists in practical recommendations.

\noindent
\textbf{Cross-domain Recommendation.}
Cross-domain recommendation attempts to learn useful knowledge from the source domain to help the target domain's recommendation. EMCDR \cite{man2017cross} is a classical embedding mapping approach, which builds the mapping function via aligned users' representations. SSCDR \cite{kang2019semi} designs a semi-supervised manner to learn item mapping based on EMCDR.
In contrast, CoNet \cite{hu2018conet} is another classical type of CDR method that uses the cross connection unit to model domain interactions.
\citet{zhao2019cross} combines items of both source and target domains in one graph to learn representations.
Dual transfer \cite{li2020ddtcdr}, source-target mixed attention \cite{ouyang2020minet}, and meta networks \cite{zhu2021transfer,zhu2022personalized} are also proposed for CDR.
Additional review information is also used to enhance the attentive knowledge transfer \cite{zhao2020catn}.
However, most of these CDR models are specially designed for the ranking module (which involve user-item interactions in cross-domain modeling).
ICAN \cite{xie2020internal} is the most related work, which captures field-level feature interactions to improve matching in multiple domains.
In CCDR, we introduce several novel CL tasks. To the best of our knowledge, we are the first to conduct CL to jointly improve representation learning and knowledge transfer in CDR.

\noindent
\textbf{Contrastive Learning.}
Contrastive learning (CL) is a representative self-supervised learning (SSL) method, which aims to learn models by contrasting positive pairs against negative pairs.
MoCo \cite{he2020momentum} builds a large dynamic dictionary with a queue and a moving-averaged momentum encoder. SimCLR \cite{chen2020simple} designs a simple contrastive learning framework with a composition of data augmentations and projectors for CL.
BYOL \cite{grill2020bootstrap} relies on its online and target networks, which iteratively bootstraps the outputs of a network to serve as targets for learning.
Some works also consider graph contrastive learning \cite{you2020graph,qiu2020gcc}.

\noindent
\textbf{CL in recommendation.}
Recently, SSL and CL are also verified in recommendation \cite{wei2021contrastive,zeng2021knowledge}.
$\mathrm{S}^3$-Rec \cite{zhou2020s3} builds contrastive learning tasks among items, attributes, sentences, and sub-sentences in sequential recommendation. UPRec focuses on user-aware SSL \cite{xiao2021uprec}. \citet{wu2022multi} adopts CL between behaviors and models in multi-behavior recommendation.
Moreover, CL has also been used in disentangled recommendation \cite{zhou2021contrastive}, social recommendation \cite{yu2021self}, sequential recommendation \cite{xie2021adversarial,wu2022personalized,wu2022selective}.
For graph-based CL, \citet{wu2021self} introduces embedding, node, edge dropouts to graph-based recommendation.
Differing from these works, we build three CL tasks to facilitate the user preference transfers between different domains in cold-start cross-domain recommendation.

\section{Methodology}

In this work, we propose CCDR to enhance the cross-domain recommendation in matching via contrastive learning.

\subsection{Problem Definition and Overall Framework}
\label{sec.overall}

\textbf{CDR in matching.}
We concentrate on the matching module of the classical two-stage recommendation systems \cite{covington2016deep}. Matching is the first step before ranking, which attempts to efficiently retrieve hundreds of items from million-level item candidates. It cares more about whether good items are retrieved (often measured by hit rate), not the specific top item ranks which should be considered by the following ranking module (often measured by NDCG or AUC) \cite{lv2019sdm,xie2020internal}. The CDR in matching task attempts to improve the target domain's matching module with the help of the source domain.

\noindent
\textbf{Overall framework.}
CCDR is trained with three types of losses, including the original source/target single-domain losses, the intra-domain CL loss, and the inter-domain CL loss.
(1) We first build a huge global diversified preference network separately for each domain as the sources of user preferences. This diversified preference network contains various objects such as user, item, tag, category, media, and word with their interactions to bring in user diverse preferences from different aspects.
(2) Next, we train the single-domain matching model via a GNN aggregator and the neighbor-similarity based loss.
(3) Since the cold-start domain lacks sufficient user behaviors, we introduce the intra-domain CL inside the target domain to train more reliable node representations with a sub-graph based data augmentation.
(4) To enhance the cross-domain knowledge transfer, we design three inter-domain CL tasks via aligned users, taxonomies, and their neighbors between two domains, which cooperate well with the diversified preference network.
All three losses are combined under a multi-task learning framework.

\subsection{Diversified Preference Network}
\label{sec.heterogeneous_network}

Conventional matching \cite{covington2016deep} and CDR \cite{man2017cross,kang2019semi} models usually heavily rely on user-item interactions to learn CTR objectives and cross-domain mapping. However, it will decrease the diversity of matching due to the popularity bias issue. Moreover, it does not take full advantage of other connections (e.g., tags, words, medias) besides users between different domains, which is particularly informative in cross-domain knowledge transfer.

Therefore, inspired by \cite{liu2020graph,xie2021improving}, we build a global diversified preference network for each domain, considering $6$ types of important objects in recommendation as nodes and their heterogeneous interactions as edges.
Specifically, we use \emph{item, user, tag, category, media, and word} as nodes. Tags and categories are item taxonomies that represent users' fine- and coarse- granularity interests. Media indicates the item's producer. Words reflect the semantic information of items extracted from items' titles or contents. To alleviate data sparsity and accelerate our offline training, we also gather users into user groups according to their basic profiles (all users having the same gender-age-location attributes are clustered in the same user group). These user groups are viewed as user nodes in CCDR.

As for edges, we consider the following six item-related interactions:
(a) \emph{User-item edge (U-I)}. This edge is generated if an item is interacted by a user group at least $3$ times. We jointly consider multiple user behaviors (i.e., click, like, share) to build this edge with different weights.
(b) \emph{Item-item edge (I-I)}. The I-I edge introduces sequential information of user behaviors in sessions. It is built if two items appear in adjacent positions in a session.
(c) \emph{Tag-item edge (T-I)}. The T-I edges connect items and their tags. It captures items' fine-grained taxonomy information.
(d) \emph{Category-item edge (C-I)}. It records items' coarse-grained taxonomy information.
(e) \emph{Media-item edge (M-I)}. It links items with their producers/sources.
(f) \emph{Word-item edge (W-I)}. It highlights the semantic information of items from their titles.
Each edge is undirected but empirically weighted according to the edge type and strength (e.g., counts for U-I edges).
Compared with conventional U-I graphs, our diversified preference network tries its best to describe items from different aspects via these heterogeneous interactions. The advantages of this diversified preference network are:
(1) it brings in additional information as supplements to user-item interactions, which jointly improve accuracy and diversity (Sec. \ref{sec.neighbor_similarity_loss}).
(2) It can build more potential bridges between different domains via users, tags, categories, and words, which cooperates well with the inter-CL tasks and the online multi-channel matching in CDR (Sec. \ref{sec.inter_CL} and Sec. \ref{sec.online_deployment}).

\subsection{Single-domain GNN Aggregator}
\label{sec.single_matching}

\subsubsection{GNN-based Aggregator}

Inspired by the great successes of GNN, we adopt GAT \cite{velivckovic2018graph} as the aggregator on the diversified preference network for simplicity and universality.
Precisely, we randomly initialize $\bm{e}^0_i$ for all heterogeneous nodes. For a node $e_i$ and its neighbor $e_k \in N_{e_i}$ ($N_{e_i}$ is the neighbor set of $e_i$ after a weighted sampling), we have $e_i$'s node representation $\bm{e}_i^x$ at the $x$-th layer as:
\begin{equation}
\begin{split}
\bm{e}_i^x = \sigma (\sum_{e_{k} \in N_{e_i}} \alpha^x_{ik} \bm{W}^{x} \bm{e}^{x-1}_{k} ).
\label{eq.aggregation}
\end{split}
\end{equation}
$\bm{W}^{x}$ is the weighting matrix, $\sigma$ is the sigmoid function. $\alpha^x_{ik}$ represents the attention between $e_i$ and $e_k$ in $x$-th layer noted as:
\begin{equation}
\begin{split}
\alpha^{x}_{ik} = \frac {\exp ( f( \bm{w}^{x^{\top}} [\bm{W}^x \bm{e}^{x-1}_i || \bm{W}^x \bm{e}^{x-1}_{k} ]))}
{\sum_{e_{l} \in N_{e_i}} \exp ( f( \bm{w}^{x^{\top}} [\bm{W}^x \bm{e}^{x-1}_i || \bm{W}^x \bm{e}^{x-1}_{l} ]))},
\label{eq.short_attention}
\end{split}
\end{equation}
where $f(\cdot)$ indicates a LeakyReLU activation and $||$ indicates the concatenation. $\bm{w}^{x}$ is the weighting vector. Note that the $N_{e_i}$ is a dynamic neighbor set which is randomly generated based on the edge weight in Sec. \ref{sec.heterogeneous_network}.
We conduct a two-layer GAT to generate the aggregated node representations $\bm{e}_i$ for all nodes ($\bm{e}^s_i$ and $\bm{e}^t_i$ for the source and target domains). It is also not difficult to conduct other GNN models such as LightGCN \cite{he2020lightgcn} in this module.

\subsubsection{Neighbor-similarity Based Optimization}
\label{sec.neighbor_similarity_loss}

\begin{figure}[!hbtp]
\centering
\includegraphics[width=0.96\columnwidth]{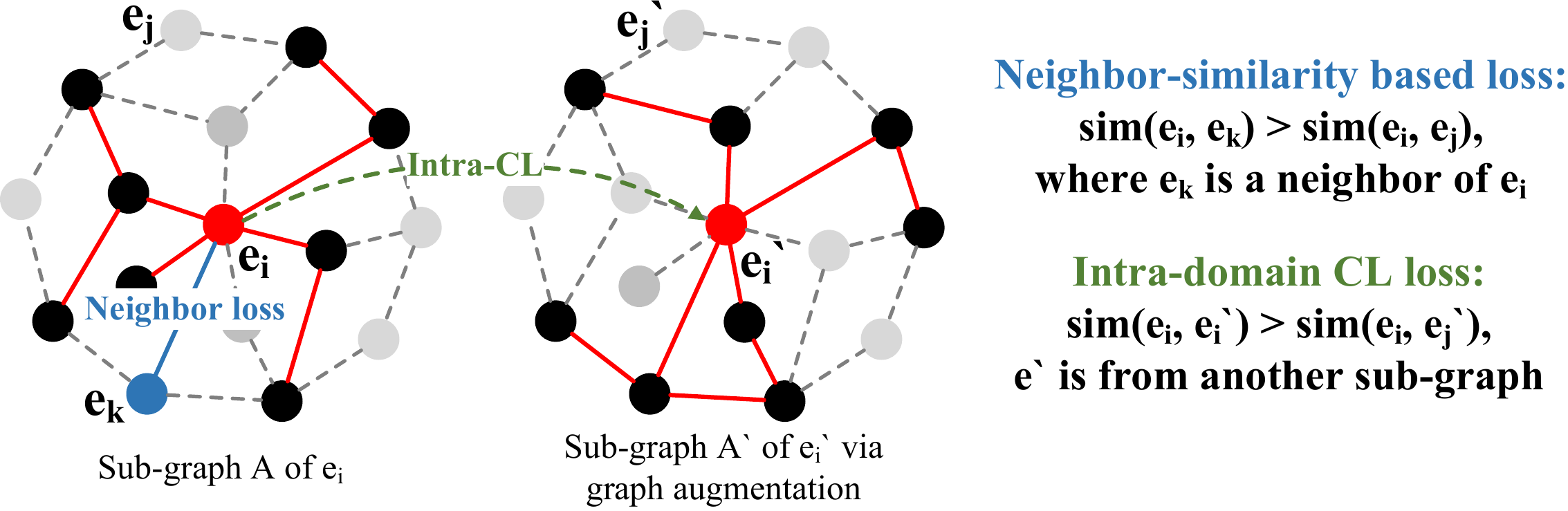}
\caption{The neighbor-similarity loss and the intra-CL loss.}
\label{fig:intra_CL}
\end{figure}

In practical CDR scenarios, users often have fewer historical behaviors on items in (cold-start) target domains. Conventional embedding-based matching methods such like Matrix factorization (MF) \cite{koren2009matrix} cannot get sufficient supervised information from the sparse user-item interactions, and thus cannot learn reliable user and item representations for matching.
To capture additional information from behavior, session, taxonomy, semantics, and data source aspects, we conduct the neighbor-similarity based loss \cite{liu2020graph} on the diversified preference network. As shown in Fig. \ref{fig:intra_CL}, this loss projects all nodes into the same latent space, making all nodes similar with their neighbors. It regards all types of edges as unsupervised information to guide the training besides user-item interactions. Formally, the neighbor-similarity based loss $L_N$ is defined as follows:
\begin{equation}
\begin{split}
L_N = -\sum_{e_i}\sum_{e_k \in N_{e_i}}\sum_{e_j \notin N_{e_i}}(- \log(\sigma({\bm{e}^\top_i}\bm{e}_j)) + \log(\sigma({\bm{e}^\top_i}\bm{e}_k))).
\end{split}
\label{eq.L_N}
\end{equation}
$\bm{e}_i$ is the $i$-th aggregated node representation, and $e_k$ is a sampled neighbor of $e_i$. $e_j$ is a randomly selected negative sample of $e_i$.

We choose the neighbor-similarity based loss for the following advantages:
(1) $L_N$ makes full use of all types of interactions between heterogeneous objects in matching, which contain significant information from user behaviors (U-I edges), sessions (I-I edges), item taxonomies (T-I and C-I edges), data sources (M-I edges) and semantics (W-I edges). It helps to capture user diverse preferences to balance accuracy and diversity in matching. If we only consider U-I edges, this loss will degrade into the classical MF.
(2) CDR in matching should deal with long-tail items. $L_N$ brings in additional information for long-tail items that can benefit cold-start domains.
(3) We conduct a cross-domain multi-channel matching strategy in online for diversity. This embedding-based retrieval strategy also depends on heterogeneous node embeddings optimized by $L_N$ to retrieve similar items in the (cold-start) target domain (see Sec. \ref{sec.online_deployment} for more details). The $L_N$ loss exactly fits the online multi-channel matching, and also well cooperates with the diversified preference network and the inter-CL losses. We cannot conduct complicated user-item interaction calculations in Eq. (\ref{eq.L_N}), since we rely on the fast embedding-based retrieval in matching for efficiency.

\subsection{Intra-domain Contrastive Learning}
\label{sec.intra_CL}

Contrastive learning is a widely-used SSL method that can make full use of unlabelled data via its pair-wise training. In CCDR, we conduct two types of CL tasks. The \textbf{intra-domain contrastive learning (intra-CL)} is conducted inside the target domain to learn better node representations, while the \textbf{inter-domain contrastive learning (inter-CL)} is adopted across the source and target domains to guide a better knowledge transfer.

In intra-CL, we conduct a sub-graph based data augmentation for each node aggregation, which could be regarded as a dynamic node/edge dropout in classical graph augmentation \cite{wu2021self}. Precisely, for a node $e_i$, we sample two neighbor set $N_{e_i}$ and  $N'_{e_i}$ to conduct the GNN aggregation, and receive two node representations $\bm{e}_i$ and $\bm{e}'_i$. $\bm{e}'_i$ is regarded as the positive instance of $\bm{e}_i$ in intra-CL, with a different sub-graph sampling focusing on different neighbors of $e_i$. Similar to \cite{chen2020simple}, we randomly sample from other examples $\bm{e}'_j$ in the same batch $B$ of $e_i$ to get the negative samples $e_j$. We do not use all examples in $B$ as negative samples for efficiency. In this case, the popularity bias is partially solved \cite{wu2019noise}. Formally, we follow the InfoNCE \cite{oord2018representation} to define the intra-CL loss $L_{intra}$ as follows:
\begin{equation}
\begin{split}
L_{intra} = - \sum_{B} \sum_{e_i \in B} \log \frac{\exp (\mathrm{sim} (\bm{e}_i,\bm{e}'_i)/\tau)}
{\sum_{{e}'_j \in S_{Bi}} \exp (\mathrm{sim} (\bm{e}_i,\bm{e}'_j)/\tau)}.
\end{split}
\label{eq.L_intra}
\end{equation}
$S_{Bi}$ indicates the negative samples of $e_i$ in $B$. $\tau$ is the temperature. $\mathrm{sim}(\bm{e}_i,\bm{e}'_j)$ measures the similarity between $\bm{e}_i$ and $\bm{e}'_j$, which is calculated as their cosine similarity.
With the intra-CL loss, long-tail nodes can also get training opportunities via SSL.

\subsection{Inter-domain Contrastive Learning}
\label{sec.inter_CL}

\begin{figure}[!hbtp]
\centering
\includegraphics[width=0.96\columnwidth]{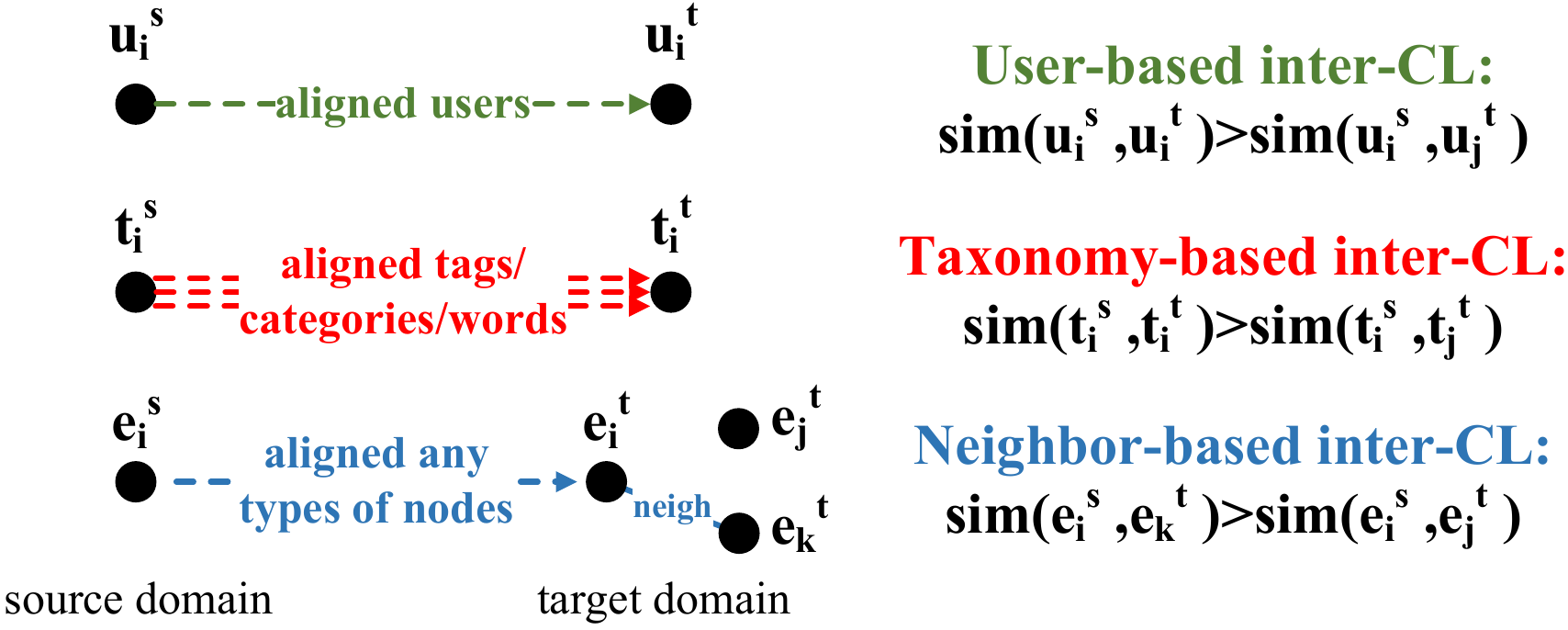}
\caption{Three inter-CL tasks across different domains.}
\label{fig:inter_CL}
\end{figure}

The inter-CL aims to improve the knowledge transfer across different domains via various types of nodes and edges in the diversified preference network. Precisely, we design three inter-domain CL tasks via aligned users, taxonomies, and neighbors as in Fig. \ref{fig:inter_CL}.

\subsubsection{User-based Inter-CL}

Most conventional CDR methods \cite{man2017cross} take aligned users as their dominating mapping seeds across domains. We follow this idea and conduct a user-based inter-CL task. Each user $u_i$ has two user representations $\bm{u}_i^s$ and $\bm{u}_i^t$ in the source and target domains learned in Sec. \ref{sec.single_matching}. Although users may have different preferences and behavior patterns in two domains, it is still natural that the source-domain representation $\bm{u}_i^s$ should be more similar with its target-domain $\bm{u}_i^t$ than any other representations $\bm{u}_j^t$. We define the user-based inter-CL loss $L_{inter_u}$ as follows:
\begin{equation}
\begin{split}
L_{inter_u} = - \sum_{u_i} \log \frac{\exp (\mathrm{sim} (\bm{u}_i^s,\bm{u}_i^t)/\tau)}
{\sum_{u_j \in S_{u_i}} \exp (\mathrm{sim} (\bm{u}_i^s,\bm{u}_j^t)/\tau)}.
\end{split}
\label{eq.L_inter_u}
\end{equation}
$S_{u_i}$ is the sampled negative set collecting from all other users except $u_i$. $\mathrm{sim}(\cdot,\cdot)$ indicates the cosine similarity.

\subsubsection{Taxonomy-based inter-CL}

Differing from some classical CDR methods \cite{kang2019semi}, CCDR builds a diversified preference network that introduces more bridges across different domains. We assume that the same tag/category/word in different domains should have the same meanings. Hence, we design a taxonomy-based inter-CL similar to the user-based CL. we take the aggregated node representation pair ($\bm{t}_i^s$, $\bm{t}_i^t$) of the same taxonomy $t_i$ in two domains as the positive pair, where $t_i$ could be tags, categories, and words. We have:
\begin{equation}
\begin{split}
L_{inter_t} = - \sum_{t_i} \log \frac{\exp (\mathrm{sim} (\bm{t}_i^s,\bm{t}_i^t)/\tau)}
{\sum_{t_j \in S_{t_i}} \exp (\mathrm{sim} (\bm{t}_i^s,\bm{t}_j^t)/\tau)},
\end{split}
\label{eq.L_inter_t}
\end{equation}
$S_{t_i}$ is the sampled negative set of $t_i$ from all other taxonomies with the same type.
We can set different temperatures for taxonomies and users if we want to sharpen the differences of some types. $L_{inter_t}$ functions as a supplement to the original user-based mapping.

\subsubsection{Neighbor-based inter-CL}

Besides the explicit alignments of users and taxonomies across domains, there are also some essential objects such as items that do not have explicit mapping. We aim to bring in more implicit cross-domain knowledge transfer paths between unaligned nodes in two domains. We suppose that similar nodes in different domains should have similar neighbors (e.g., similar items may have similar users, taxonomies, and producers). Hence, we propose a neighbor-based inter-CL, which builds indirect (multi-hop) connections between objects in different domains. Precisely, we define $E_A$ as the overall aligned node set (including users, tags, categories, and words). The neighbor-based inter-CL loss $L_{inter_n}$ is formalized with all aligned nodes $e_i \in E_A$ and $e_i$'s neighbor set $N^t_{e_i}$ in the target domain as follows:
\begin{equation}
\begin{split}
L_{inter_n} = - \sum_{e_i \in E_A} \sum_{e_k \in N_{e_i}^t} \log \frac{\exp (\mathrm{sim} (\bm{e}_i^s,\bm{e}_k^t)/\tau)}
{\sum_{e_j \notin N_{e_i}^t} \exp (\mathrm{sim} (\bm{e}_i^s,\bm{e}_j^t)/\tau)}.
\end{split}
\label{eq.L_inter_n}
\end{equation}
In $L_{inter_n}$, for an aligned node's representation $\bm{e}_i^s$ in the source domain, its target-domain neighbor's representation $\bm{e}_k^t$ is the positive instance, while other target-domain representations $\bm{e}_j^t$ are negative. It is reasonable since related objects should be connected in the diversified preference network and learned to be similar under the neighbor-similarity based loss in Eq. (\ref{eq.L_N}). It is also convenient to extend the current positive samples $e_k \in N_{e_i}^t$ to multi-hop neighbors for better generalization and diversity in CDR.

This neighbor-based inter-CL greatly multiplies the diversified knowledge transfer paths between two domains, especially for the cold-start items. For example, through the $tag_i^s \rightarrow tag_i^t \rightarrow item_j^t$ path, the cold-start $item_j$'s representation in the target domain can be directly influenced by fully-trained representations in the source domain. Moreover, the similarities between different types of source-domain node representations and the target-domain item representations are directly used in the online multi-channel matching for diversified retrieval, which will be discussed in Sec. \ref{sec.online_deployment}.

Finally, we combine all three CL losses from aligned user, taxonomy, and neighbor aspects to form the inter-CL loss $L_{inter}$ as:
\begin{equation}
\begin{split}
L_{inter} = L_{inter_u} + L_{inter_t} + L_{inter_n}.
\end{split}
\label{eq.L_inter}
\end{equation}

\subsection{Multi-task Optimization}
\label{sec.multi_task_optimization}

Following classical CL-based recommendation models \cite{yu2021self}, we also conduct a multi-task optimization combining the source-domain matching loss $L_{N_s}$, the target-domain matching loss $L_{N_t}$, the intra-CL loss $L_{intra}$, and the inter-CL loss $L_{inter}$ as follows:
\begin{equation}
\begin{split}
L = \lambda_1 L_{N_s} + \lambda_2 L_{N_t} + \lambda_3 L_{intra} + \lambda_4 L_{inter}.
\end{split}
\label{eq.L_all}
\end{equation}
$\lambda_1, \lambda_2, \lambda_3, \lambda_4$ are loss weights set as $1.0, 1.0, 1.5, 0.6$ according to the grid search.
More details and motivations are in Appendix \ref{app:implementation}.

\section{Online Deployment}
\label{sec.online_deployment}

We have deployed CCDR on the cold-start matching module in a well-known recommendation system named WeChat Top Stories.
A good CDR-based cold-start matching module should have the following key characteristics: (1) making full use of user behaviors and item features in the source and target domains, (2) capturing user diverse preferences from different aspects, and (3) balancing accuracy, diversity and efficiency. To achieve these, we propose a new \textbf{cross-domain multi-channel matching} in online.

Specifically, we conduct six channels including \emph{user}, \emph{item}, \emph{tag}, \emph{category}, \emph{media}, and \emph{word} channels to retrieve items in the target domains via node representations learned by Eq. (\ref{eq.L_all}). We rely on the user historical behavior sequence $seq=\{d_1, d_2, \cdots, d_n\}$ to capture user's interests, where $d_i$ is the $i$-th clicked item and $n$ is the max length. In the item channel, we directly use the node representations of all items in $seq$ to retrieve similar items in the target domain. Formally, we define the score $s_i^d$ of the $i$-th target-domain item $\bar{d}_i$ in the item channel as follows:
\begin{equation}
\begin{split}
s_i^d=\sum_{j=1} sim(\bar{\bm{d}}_i,\bm{d}_j) \times satisf_j \times recency_j \times z^d(i,j).
\end{split}
\label{eq.item_channel}
\end{equation}
$sim(\bar{\bm{d}}_i,\bm{d}_j)$ is the cosine similarity between the clicked item $d_j$ in user historical behaviors and the item candidate $\bar{d}_i$ in the target domain, where $\bar{\bm{d}}_i$ and $\bm{d}_j$ are aggregated item embeddings trained by Eq. (\ref{eq.L_all}). $satisf_j$ measures the posterior user satisfaction on $d_j$, which is calculated as user's complete rate on the item.
$recency_j$ models the temporal factors of historical items, which decays exponentially from the short term to the long term ($recency_j=0.95^{n-j}$).
For online efficiency, each item in $seq$ only recommends its top $100$ nearest items. $z^d(i,j)$ equals $1$ only if the target-domain item $\bar{d}_i$ appears in the top $100$ nearest items of $d_j$, and otherwise $z^d(i,j)=0$. We pre-calculate the similarities and index the top nearest items for all nodes in offline to further accelerate the online matching.

To capture user diverse preferences from different aspects, we further conduct the tag, category, media and word channels similar to the item channel. Taking the tag channel for instance, we build a historical tag sequence $seq_t=\{T_1, T_2, \cdots, T_n\}$ according to the item sequence $\{d_1, d_2, \cdots, d_n\}$, where $T_j$ is the tag set of $d_j$. All tags in $seq_t$ retrieve top $100$ nearest items in the target domains as candidates. Similar to Eq. (\ref{eq.item_channel}), the score of the $i$-th target-domain item $\bar{d}_i$ in the tag channel $s_i^t$ is defined as follows:
\begin{equation}
\begin{split}
s_i^t=\sum_{j=1}^{n} \sum_{t_k} sim(\bm{\bar{d}}_i,\bm{t}_k) \times satisf_j \times recency_j \times z^t(i,j,k).
\end{split}
\end{equation}
$sim(\bm{\bar{d}}_i,\bm{t}_k)$ is the cosine similarity between $\bm{\bar{d}}_i$ and the aggregated tag representation $\bm{t}_k$. $z^t(i,j,k)$ is the tag's indicator. $z^t(i,j,k)=1$ only if the tag $t_k$ belongs to the $j$-th item $d_j$ in $seq$, and $\bar{d}_i$ locates in the top $100$ nearest items of $t_k$. Other category, media and word channels are the same as the tag channel, generating their corresponding scores $s_i^c$, $s_i^m$ and $s_i^w$.
As for the user channel, we directly depend on the user's gender-age-location attribute triplet's (i.e., the user group in Sec. \ref{sec.heterogeneous_network}) node representations to retrieve top nearest items according to the cosine similarity score $s_i^u$ for $\bar{d}_i$.

Finally, all top items retrieved by six heterogeneous channels are combined and reranked via the aggregated score $s_i$ as:
\begin{equation}
\begin{split}
s_i=s_i^u+s_i^d+s_i^t+s_i^c+s_i^m+s_i^w.
\end{split}
\label{eq.all_score}
\end{equation}
It is easy to set and adjust the hyper-parameters of heterogeneous channels' weights for the practical demands and the preferences of systems.
We rank top target-domain items via $s_i$, and select top $500$ items as the final output of our multi-channel matching, considering both matching accuracy and memory/computation costs.

We conclude the feasibility and advantages of our cross-domain multi-channel matching as follows:
(1) These multiple matching channels rely on the similarities between the target-domain items and heterogeneous nodes, which is consistent with the neighbor-similarity based loss and the inter-CL losses.
(2) The multi-channel matching makes full use of all heterogeneous information to generate diversified item candidates, which is essential in cold-start matching.
(3) We pre-calculate the indexes for the top nearest items of all nodes, which greatly reduces the online computation costs. The online computation complexity of CCDR is $O(\log(600n)+600n)$ ($n$ is the length of user historical behavior).
More details of online deployment and efficiency are given in Appendix \ref{app:online_deployment}.

\section{Experiments}

\subsection{Large-scale CDR Matching Dataset}
\label{sec.dataset}

CCDR relies on item-related taxonomy, semantic, and producer information for CDR in matching, while no large-scale public CDR dataset is capable for this setting. Therefore, we build a new CDR dataset CDR-427M extracted from a real-world recommendation system named WeChat Top Stories, which contains a source domain and two target domains. Specifically, we randomly select nearly $63$ million users, and collect their $427$ million behaviors on $3.0$ million items. We split these behaviors into the train set and the test set using the chronological order. We also bring in $187$ thousand tags, $356$ categories, $56$ thousand medias, and $207$ thousand words as additional item-related information. All data are preprocessed via data masking to protect the user's privacy.

To simulate different CDR scenarios,  we evaluate on two target domains having different cold-start degrees. The first is a few-shot target domain, where most users only have several behaviors. The second is a strict cold-start domain, which is more challenging since all user behaviors on items in the train set are discarded \cite{qian2020attribute}.
Following Sec. \ref{sec.heterogeneous_network}, we build three diversified preference networks for all domains separately via the train set and all item-related information. More details are in Table \ref{tab:dataset} and Appendix \ref{app:dataset}.

\begin{table}[!hbtp]
\centering
\small
\caption{Statistics of three domains in CDR-427M.}
\begin{tabular}{l|ccccc}
\toprule
domain & user & item & edge & train & test\\
\midrule
source domain & 63.1M & 2.04M & 175M & 406M & / \\
few-shot target domain & 2.38M& 0.39M & 29.8M & 10.8M & 4.99M \\
strict cold-start domain & 2.23M & 0.57M & 15.0M & / & 5.48M \\
\bottomrule
\end{tabular}
\label{tab:dataset}
\end{table}

\begin{table*}[!hbtp]
\caption{Results of matching-related metrics on CDR-427M. All improvements of CCDR are significant (t-test with $p < 0.01$). * indicates that these models are based on the same single-domain model (noted as GraphDR+) in CCDR.}
\centering
\small
\begin{tabular}{l||
p{1.1cm}<{\centering}|p{1.1cm}<{\centering}|p{1.1cm}<{\centering}|p{1.1cm}<{\centering}||
p{1.1cm}<{\centering}|p{1.1cm}<{\centering}|p{1.1cm}<{\centering}|p{1.1cm}<{\centering}}
\toprule
\multirow{2}{*}{Model} & \multicolumn{4}{c||}{few-shot target domain} & \multicolumn{4}{c}{strict cold-start target domain} \\
\cmidrule{2-9}
 & HIT@50 & HIT@100 & HIT@200 & HIT@500 & HIT@50 & HIT@100 & HIT@200 & HIT@500 \\
\midrule
FM (\citeauthor{rendle2010factorization} \citeyear{rendle2010factorization}) & 0.0101 & 0.0176 & 0.0293 & 0.0551 & 0.0048 & 0.0084 & 0.0145 & 0.0299 \\
AutoInt (\citeauthor{song2019autoint} \citeyear{song2019autoint}) & 0.0125 & 0.0219 & 0.0370 & 0.0705 & 0.0058 & 0.0102 & 0.0173 & 0.0356 \\
GraphDR+ (\citeauthor{xie2021improving} \citeyear{xie2021improving})* & 0.0132 & 0.0236 & 0.0402 & 0.0780 & {0.0087} & {0.0156} & {0.0266} & {0.0543} \\
\midrule
EMCDR+ (\citeauthor{man2017cross} \citeyear{man2017cross})* & 0.0137 & 0.0242 & 0.0414 & 0.0801 & 0.0091 & 0.0160 & 0.0274 & 0.0561 \\
SSCDR+ (\citeauthor{kang2019semi} \citeyear{kang2019semi})* & 0.0146 & 0.0261 & 0.0448 & 0.0879 & {0.0104} & {0.0183} & {0.0312} & {0.0637} \\
ICAN (\citeauthor{xie2020internal} \citeyear{xie2020internal}) & {0.0163} & {0.0291} & {0.0502} & {0.0994} & 0.0074 & 0.0131 & 0.0226 & 0.0464 \\
\midrule
Sub-graph CL (GraphDR+)* & 0.0174 & 0.0312 & 0.0541 & 0.1099 & 0.0118 & 0.0209 & 0.0361 & 0.0732  \\
Cross-domain KD (GraphDR+)* & 0.0169 & 0.0305 & 0.0530 & 0.1058 & 0.0131 & 0.0229 & 0.0387 & 0.0780 \\
\midrule
CCDR (inter-CL)* & 0.0213 & 0.0378 & 0.0671 & 0.1327 & 0.0154 & 0.0271 & 0.0467 & 0.0931  \\
CCDR (inter-CL+intra-CL)* & \textbf{0.0225} & \textbf{0.0411} & \textbf{0.0715} & \textbf{0.1442} & \textbf{0.0161} & \textbf{0.0284} & \textbf{0.0487} & \textbf{0.0987} \\
\bottomrule
\end{tabular}
\label{tab:main_results}
\end{table*}

\subsection{Competitors}
\label{sec.baseline}

We implement several competitive baselines focusing on the matching module and cross-domain recommendation for comparisons.

\paragraph{Classical Matching Methods}

We implement three competitive matching models as baselines. We do not compare with tree-based models \cite{zhu2018learning}, for they cannot be deployed in cold-start domains. All user behaviors of two domains are considered in these models.

\begin{itemize}
 \item \textbf{Factorization Machine (FM) [\citeauthor{rendle2010factorization} \citeyear{rendle2010factorization}].} FM \cite{rendle2010factorization} is a classical embedding-based matching model. It captures the feature interactions between users and items for embedding-based retrieval under the two-tower architecture \cite{covington2016deep}.
 \item \textbf{AutoInt  [\citeauthor{song2019autoint} \citeyear{song2019autoint}].} AutoInt \cite{song2019autoint} is a recent method that utilizes self-attention to model feature interactions. It also adopt the two-tower architecture for matching.
 \item \textbf{GraphDR+  [\citeauthor{xie2021improving} \citeyear{xie2021improving}].} GraphDR \cite{xie2021improving} is an effective graph-based matching model. The single-domain model of CCDR could be considered as an enhanced GraphDR with differences in node aggregation and multi-channel matching specially designed for CDR. We directly conduct the single-domain model of CCDR on the joint network containing both source and target domains, noted as GraphDR+.
\end{itemize}

\paragraph{Cross-domain/Multi-domain Methods}

We also implement two representative CDR models and one multi-domain matching model as baselines. We do not compare with CDR models like CoNet \cite{hu2018conet}, since they cannot be directly used in matching for efficiency.

\begin{itemize}
 \item \textbf{EMCDR+ [\citeauthor{man2017cross} \citeyear{man2017cross}].} EMCDR \cite{man2017cross} is a classical CDR model that directly learns the embedding mapping of users between two domains. For fair comparisons, we use the same single-domain model and multi-channel matching in CCDR for learning and serving, noted as EMCDR+.
 \item \textbf{SSCDR+ [\citeauthor{kang2019semi} \citeyear{kang2019semi}].} SSCDR \cite{kang2019semi} is regarded as an enhanced EMCDR, which adopts a semi-supervised loss to learn the mapping of items. We also follow the same settings of EMCDR to get SSCDR+. Since the strict cold-start domain has no user-item behaviors, we use aligned taxonomies to learn cross-domain mappings in EMCDR+ and SSCDR+.
 \item \textbf{ICAN [\citeauthor{xie2020internal} \citeyear{xie2020internal}].} ICAN \cite{xie2020internal} is the SOTA model in multi-domain matching, which is the most related work of our task. It highlights the interactions between feature fields in different domains for cold-start matching.
\end{itemize}

\paragraph{Knowledge Distillation/Contrastive Learning Methods}

We further propose two enhanced versions of the single-domain matching model in CCDR (i.e., GraphDR+) for more challenging comparisons.

\begin{itemize}
 \item \textbf{Sub-graph CL.}
 We build a sub-graph CL method based on GraphDR+. It considers the intra-CL loss with a sub-graph augmentation in Eq. (\ref{eq.L_intra}) inspired by \cite{qiu2020gcc,wu2021self}. It can be viewed as an ablation version of CCDR without the inter-CL.
 \item \textbf{Cross-domain KD.} We further propose a cross-domain knowledge distillation (KD) model. This model also follows the single-domain model of CCDR, learning the cross-domain mapping via the Hint loss \cite{romero2015fitnets} between aligned nodes in two domains (i.e., user, tag, category, and word).
\end{itemize}

\subsection{Experimental Settings}
\label{sec.experimental_settings}

In the single-domain model of CCDR, the input dimensions of all nodes are $128$, and the output dimensions are $100$. We conduct a weighted neighbor sampling to select $25$ and $10$ neighbors for the first and second layers' aggregations. The edge weight is proportional to the mutual information between its two nodes to make sure that different types of interactions can have comparable frequencies.
In online matching, we use the top $200$ most recent behaviors. All graph-based models have the same online matching strategy. The batch sizes and the negative sample numbers of the intra-CL, inter-CL, and neighbor-similarity based losses are $4,096$ and $10$. The temperature $\tau$ is set to be $1$.
For all models, we conduct a grid search to select parameters.
Parameter analyses of CL loss weights are given in Sec. \ref{sec.model_analyses}.
All models share the same experimental settings and multi-domain behaviors for fair comparisons.

\subsection{Evaluation of CDR in Matching (RQ1)}
\label{sec.offline_evaluation}

\subsubsection{Evaluation Protocols}

We evaluate on the few-shot and strict cold-start domains separately. All models select top $N$ items from the overall corpora for each test instance. Following classical matching models \cite{sun2019bert4rec,xie2020internal,xie2021improving}, we utilize the top $N$ hit rate (HIT@N) as our evaluation metric. To simulate the real-world matching systems, we concentrate on larger $N$ as $50$, $100$, $200$, and $500$ (we retrieve top $500$ items in online).
We should double clarify that CCDR focuses on CDR in matching, which \textbf{\emph{cares whether good items are retrieved}}, not the specific ranks that should be measured by ranking. Hence, HIT@N is much more suitable for matching than ranking metrics such as AUC and NDCG.
We also evaluate the diversity via a classical aggregate diversity metric named item coverage \cite{herlocker2004evaluating}.

\subsubsection{Experimental Results}

From Table \ref{tab:main_results} we can observe that:

(1) CCDR achieves significant improvements over all baselines on all HIT@N in both two domains, with the significance level $p<0.01$ (the deviations of CCDR are within $\pm0.0004$ in HIT@500). It indicates that CCDR can learn high-quality matching embeddings and well transfer useful knowledge to the target domain via CL. The improvements of CCDR mainly derive from three aspects:
(a) The intra-CL enables more sufficient and balanced training via SSL with selected negative samples, which successfully alleviates the data sparsity and popularity bias issues.
(b) The inter-CL builds interactions across different domains via three CL tasks, which multiplies the knowledge transfer via heterogeneous bridges.
(c) The diversified preference network, CL losses, and multi-channel matching cooperate well with each other. The similarities used in online matching are directly optimized via losses in Eq. (\ref{eq.L_all}).

(2) CCDR has large improvement percentages on the challenging strict cold-start domain ($55\%$ improvement on HIT@500 over SSCDR+), where users have no behaviors on target items. It is natural since the combination of the diversified preference network and user/taxonomy/neighbor based inter-CL tasks can transfer more diversified preferences via more cross-domain bridges.
Moreover, comparing with different CCDR models, we find that both intra-CL and inter-CL are effective, while inter-CL plays a more important role in CDR.
We also find that CCDR has $4.2\%$ and $6.0\%$ improvements on the diversity metric \emph{item coverage} \cite{herlocker2004evaluating} compared to the best-performing GraphDR+ in two domains. It indicates that CCDR has better performances on the diversity via CL tasks.

(3) Among baselines, we find that ICAN performs better in the few-shot domain, while SSCDR+ performs better in the strict cold-start domain. It is because that ICAN strongly relies on the feature field interactions between behaviors in different domains, which are extremely sparse or even missing in the strict cold-start scenarios. In contrast, SSCDR+ benefits from cross-domain mapping.
Moreover, classical matching methods such as GraphDR+ perform worse than CCDR. It implies that simply mixing behaviors in two domains may not get good performances, since the unbalanced domain data will confuse the user preference learning in the target domain.

(4) Comparing with different CDR methods, we observe that the CL-based methods are the most effective compared to knowledge distillation (e.g., cross-domain KD) and embedding mapping (e.g., EMCDR+ and SSCDR+). It is because that (a) contrastive learning can provide a sufficient and balanced training via SSL, and (b) CCDR conducts knowledge transfer via not only aligned users, but also taxonomies and neighbors. In this case, the popularity bias and data sparsity issues in the CDR part can be largely alleviated.

\subsection{Online Evaluation (RQ2)}
\label{sec.online_evaluation}

\subsubsection{Evaluation Protocols}

To verify the effectiveness of CCDR in real-world scenarios, we conduct an online A/B test on a well-known online recommendation system named WeChat Top Stories. Precisely, we deploy CCDR and several competitive baselines in the matching module of a relatively cold-start domain as in Sec. \ref{sec.online_deployment}, with the ranking module unchanged. The online baseline is the GraphDR+ (target) model trained solely on the target domain. In online evaluation, we focus on three online metrics in the target domain: (1) CTR, (2) average user duration per capita, and (3) average share rate per capita. We conduct the A/B test for $8$ days.

\begin{table}[!hbtp]
\centering
\small
\begin{tabular}{l|ccc}
\toprule
Model & CTR & Duration & Share rate \\
\midrule
GraphDR+ (source+target) & +1.121\% & +2.096\% & +1.416\%  \\
Sub-graph CL & +2.182\% & +2.213\% & +2.378\%  \\
Cross-domain KD & +2.056\% & +2.959\% & +6.244\%  \\
CCDR (inter-CL) & +3.375\% & +4.354\% & +7.947\%  \\
CCDR (inter-CL+intra-CL) & \textbf{+14.368\%} & \textbf{+6.623\%} & \textbf{+10.401\%}  \\
\bottomrule
\end{tabular}
\caption{Online A/B tests on WeChat Top Stories.}
\label{tab:online}
\end{table}

\subsubsection{Experimental Results}

Table \ref{tab:online} shows the online improvement percentages of all models. We can find that:

(1) CCDR significantly outperforms all models in three metrics with the significance level $p<0.01$. Note that all models are based on the same single-domain model in CCDR (i.e., GraphDR+). It reconfirms the effectiveness of the intra-domain and inter-domain contrastive learning. We jointly consider multiple behaviors such as click, share and like to build the diversified preference network, and use a neighbor-similarity based loss to learn user diverse preferences. Hence, CCDR has improvements on different metrics, which reflects user's real satisfaction more comprehensively.

(2) Comparing with the base model that only considers the target domain, we know that the source domain's information is essential. Looking into the differences among GraphDR+, Sub-graph CL (i.e., CCDR (intra-CL)), and CCDR (inter-CL), we can find that both intra-CL and inter-CL are effective in online scenarios. Moreover, CCDR models outperform GraphDR+ and Cross-domain KD, which also verifies the advantages of inter-CL over simple multi-domain mixing and cross-domain knowledge distillation in knowledge transfer.

\subsection{Ablation Study (RQ3)}
\label{sec.ablation}

In this subsection, we further compare CCDR with its several ablation versions to show the effectiveness of different CL tasks. Table \ref{tab:ablation_test} displays the HIT@N results on both few-shot and strict cold-start domains. We find that:
(1) Both intra-CL and inter-CL are essential in few-shot and cold-start domains. Inter-CL contributes the most to the CDR performances, since it is strongly related to the knowledge transfer task in CDR and fits well with the neighbor-similarity based loss of the single-domain model.
(2) The intra-CL task also significantly improves the matching in CDR, while it just achieves slight improvements on the strict cold-start domain. The power of intra-CL will be multiplied when there are more user behaviors in the target domain.
(3) From the second ablation group, we observe that all three inter-CL tasks can provide useful information for CDR. We observe that the user-based inter-CL functions well in the few-shot domain (since it has more user-related interactions), while taxonomy-based inter-CL achieves higher improvements in the cold-start domain. Note that CCDR does not use the user channel and adopt $L_{inter_u}$ in the strict cold-start domain, since the cold-start user nodes are isolated in the target domain with no behaviors.

\begin{table}[!hbtp]
\centering
\small
\begin{tabular}{l|c|c|c|c}
\toprule
\multirow{2}{*}{Model} & \multicolumn{2}{c|}{few-shot} & \multicolumn{2}{c}{strict cold-start} \\
\cmidrule{2-5}
 & HIT@50 & HIT@500 & HIT@50 & HIT@500 \\
\midrule
CCDR & 0.0225 & 0.1442 & 0.0161 & 0.0987 \\
\midrule
  \quad w/o intra-CL  & 0.0213 & 0.1327 & 0.0154 & 0.0931 \\
  \quad w/o inter-CL  & 0.0174 & 0.1099 & 0.0118 & 0.0732 \\
\midrule
  \quad w/o $L_{inter_u}$  & 0.0211 & 0.1376 & N/A & N/A \\
  \quad w/o $L_{inter_t}$  & 0.0219 & 0.1403 & 0.0149 & 0.0901 \\
  \quad w/o $L_{inter_n}$  & 0.0221 & 0.1411 & 0.0157 & 0.0948 \\
\bottomrule
\end{tabular}
\caption{Ablation tests on CDR-427M.}
\label{tab:ablation_test}
\end{table}

\subsection{Model Analyses (RQ4)}
\label{sec.model_analyses}

\begin{figure}[!hbtp]
\centering
\includegraphics[width=0.96\columnwidth]{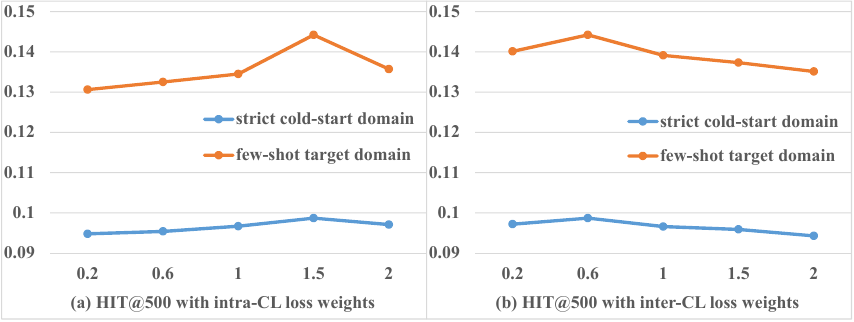}
\caption{Results of different intra-/inter- CL loss weights.}
\label{fig:paramater}
\end{figure}

We further conduct two model analyses on different weights of the intra-CL and inter-CL losses. Fig. \ref{fig:paramater} displays the HIT@500 results of different intra- and inter- CL weights ($\lambda_3$ and $\lambda_4$) on both few-shot and strict cold-start domains. We can find that:
(1) as the loss weight increases, the HIT@500 results of both intra-CL and inter-CL losses first increase and then decrease. The best parameters are $\lambda_3=1.5, \lambda_4=0.6$. Note that the parameter analysis is carried out around the optimal parameter point.
(2) The performance trends of two CL loss weights are relatively consistent on both few-shot and strict cold-start domains. Moreover, CCDR models with different CL loss weights still outperform all baselines, which verifies the robustness and usability of CCDR in real-world scenarios.

\section{Conclusion and Future Work}

In this work, we propose a novel CCDR framework to deal with CDR in matching. We adopt the intra-CL to alleviate the data sparsity and popularity bias issues in matching, and design three inter-CL tasks to enable more diverse and effective knowledge transfer. CCDR achieves significant offline and online improvements on different scenarios, and is deployed on real-world systems.
In the future, we will explore more sophisticated inter-CL tasks to further improve the effectiveness and diversity of knowledge transfer.
We will also extend inter-CL to other CDR models and knowledge transfer tasks.

\bibliographystyle{ACM-Reference-Format}
\bibliography{reference}


\begin{thebibliography}{48}


\ifx \showCODEN    \undefined \def \showCODEN     #1{\unskip}     \fi
\ifx \showDOI      \undefined \def \showDOI       #1{#1}\fi
\ifx \showISBNx    \undefined \def \showISBNx     #1{\unskip}     \fi
\ifx \showISBNxiii \undefined \def \showISBNxiii  #1{\unskip}     \fi
\ifx \showISSN     \undefined \def \showISSN      #1{\unskip}     \fi
\ifx \showLCCN     \undefined \def \showLCCN      #1{\unskip}     \fi
\ifx \shownote     \undefined \def \shownote      #1{#1}          \fi
\ifx \showarticletitle \undefined \def \showarticletitle #1{#1}   \fi
\ifx \showURL      \undefined \def \showURL       {\relax}        \fi
\providecommand\bibfield[2]{#2}
\providecommand\bibinfo[2]{#2}
\providecommand\natexlab[1]{#1}
\providecommand\showeprint[2][]{arXiv:#2}

\bibitem[Cen et~al\mbox{.}(2020)]%
        {cen2020controllable}
\bibfield{author}{\bibinfo{person}{Yukuo Cen}, \bibinfo{person}{Jianwei Zhang},
  \bibinfo{person}{Xu Zou}, \bibinfo{person}{Chang Zhou},
  \bibinfo{person}{Hongxia Yang}, {and} \bibinfo{person}{Jie Tang}.}
  \bibinfo{year}{2020}\natexlab{}.
\newblock \showarticletitle{Controllable multi-interest framework for
  recommendation}. In \bibinfo{booktitle}{\emph{Proceedings of KDD}}.
\newblock


\bibitem[Chen et~al\mbox{.}(2020)]%
        {chen2020simple}
\bibfield{author}{\bibinfo{person}{Ting Chen}, \bibinfo{person}{Simon
  Kornblith}, \bibinfo{person}{Mohammad Norouzi}, {and}
  \bibinfo{person}{Geoffrey Hinton}.} \bibinfo{year}{2020}\natexlab{}.
\newblock \showarticletitle{A simple framework for contrastive learning of
  visual representations}. In \bibinfo{booktitle}{\emph{Proceedings of ICML}}.
\newblock


\bibitem[Covington et~al\mbox{.}(2016)]%
        {covington2016deep}
\bibfield{author}{\bibinfo{person}{Paul Covington}, \bibinfo{person}{Jay
  Adams}, {and} \bibinfo{person}{Emre Sargin}.}
  \bibinfo{year}{2016}\natexlab{}.
\newblock \showarticletitle{Deep neural networks for youtube recommendations}.
  In \bibinfo{booktitle}{\emph{Proceedings of RecSys}}.
\newblock


\bibitem[Grill et~al\mbox{.}(2020)]%
        {grill2020bootstrap}
\bibfield{author}{\bibinfo{person}{Jean-Bastien Grill},
  \bibinfo{person}{Florian Strub}, \bibinfo{person}{Florent Altch{\'e}},
  \bibinfo{person}{Corentin Tallec}, \bibinfo{person}{Pierre~H Richemond},
  \bibinfo{person}{Elena Buchatskaya}, \bibinfo{person}{Carl Doersch},
  \bibinfo{person}{Bernardo~Avila Pires}, \bibinfo{person}{Zhaohan~Daniel Guo},
  \bibinfo{person}{Mohammad~Gheshlaghi Azar}, {et~al\mbox{.}}}
  \bibinfo{year}{2020}\natexlab{}.
\newblock \showarticletitle{Bootstrap your own latent: A new approach to
  self-supervised learning}.
\newblock \bibinfo{journal}{\emph{arXiv:2006.07733}} (\bibinfo{year}{2020}).
\newblock


\bibitem[Hao et~al\mbox{.}(2021)]%
        {hao2021adversarial}
\bibfield{author}{\bibinfo{person}{Xiaobo Hao}, \bibinfo{person}{Yudan Liu},
  \bibinfo{person}{Ruobing Xie}, \bibinfo{person}{Kaikai Ge},
  \bibinfo{person}{Linyao Tang}, \bibinfo{person}{Xu Zhang}, {and}
  \bibinfo{person}{Leyu Lin}.} \bibinfo{year}{2021}\natexlab{}.
\newblock \showarticletitle{Adversarial Feature Translation for Multi-domain
  Recommendation}. In \bibinfo{booktitle}{\emph{Proceedings of KDD}}.
\newblock


\bibitem[He et~al\mbox{.}(2020b)]%
        {he2020momentum}
\bibfield{author}{\bibinfo{person}{Kaiming He}, \bibinfo{person}{Haoqi Fan},
  \bibinfo{person}{Yuxin Wu}, \bibinfo{person}{Saining Xie}, {and}
  \bibinfo{person}{Ross Girshick}.} \bibinfo{year}{2020}\natexlab{b}.
\newblock \showarticletitle{Momentum contrast for unsupervised visual
  representation learning}. In \bibinfo{booktitle}{\emph{Proceedings of CVPR}}.
\newblock


\bibitem[He et~al\mbox{.}(2020a)]%
        {he2020lightgcn}
\bibfield{author}{\bibinfo{person}{Xiangnan He}, \bibinfo{person}{Kuan Deng},
  \bibinfo{person}{Xiang Wang}, \bibinfo{person}{Yan Li},
  \bibinfo{person}{Yongdong Zhang}, {and} \bibinfo{person}{Meng Wang}.}
  \bibinfo{year}{2020}\natexlab{a}.
\newblock \showarticletitle{Lightgcn: Simplifying and powering graph
  convolution network for recommendation}. In
  \bibinfo{booktitle}{\emph{Proceedings of SIGIR}}.
\newblock


\bibitem[Herlocker et~al\mbox{.}(2004)]%
        {herlocker2004evaluating}
\bibfield{author}{\bibinfo{person}{Jonathan~L Herlocker},
  \bibinfo{person}{Joseph~A Konstan}, \bibinfo{person}{Loren~G Terveen}, {and}
  \bibinfo{person}{John~T Riedl}.} \bibinfo{year}{2004}\natexlab{}.
\newblock \showarticletitle{Evaluating collaborative filtering recommender
  systems}.
\newblock \bibinfo{journal}{\emph{TOIS}} (\bibinfo{year}{2004}).
\newblock


\bibitem[Hu et~al\mbox{.}(2018)]%
        {hu2018conet}
\bibfield{author}{\bibinfo{person}{Guangneng Hu}, \bibinfo{person}{Yu Zhang},
  {and} \bibinfo{person}{Qiang Yang}.} \bibinfo{year}{2018}\natexlab{}.
\newblock \showarticletitle{Conet: Collaborative cross networks for
  cross-domain recommendation}. In \bibinfo{booktitle}{\emph{Proceedings of
  CIKM}}.
\newblock


\bibitem[Huang et~al\mbox{.}(2020)]%
        {huang2020embedding}
\bibfield{author}{\bibinfo{person}{Jui-Ting Huang}, \bibinfo{person}{Ashish
  Sharma}, \bibinfo{person}{Shuying Sun}, \bibinfo{person}{Li Xia},
  \bibinfo{person}{David Zhang}, \bibinfo{person}{Philip Pronin},
  \bibinfo{person}{Janani Padmanabhan}, \bibinfo{person}{Giuseppe Ottaviano},
  {and} \bibinfo{person}{Linjun Yang}.} \bibinfo{year}{2020}\natexlab{}.
\newblock \showarticletitle{Embedding-based retrieval in facebook search}. In
  \bibinfo{booktitle}{\emph{Proceedings of KDD}}.
\newblock


\bibitem[Johnson et~al\mbox{.}(2019)]%
        {johnson2019billion}
\bibfield{author}{\bibinfo{person}{Jeff Johnson}, \bibinfo{person}{Matthijs
  Douze}, {and} \bibinfo{person}{Herv{\'e} J{\'e}gou}.}
  \bibinfo{year}{2019}\natexlab{}.
\newblock \showarticletitle{Billion-scale similarity search with GPUs}.
\newblock \bibinfo{journal}{\emph{IEEE Trans. Big Data}}
  (\bibinfo{year}{2019}).
\newblock


\bibitem[Kang et~al\mbox{.}(2019)]%
        {kang2019semi}
\bibfield{author}{\bibinfo{person}{SeongKu Kang}, \bibinfo{person}{Junyoung
  Hwang}, \bibinfo{person}{Dongha Lee}, {and} \bibinfo{person}{Hwanjo Yu}.}
  \bibinfo{year}{2019}\natexlab{}.
\newblock \showarticletitle{Semi-supervised learning for cross-domain
  recommendation to cold-start users}. In \bibinfo{booktitle}{\emph{Proceedings
  of CIKM}}.
\newblock


\bibitem[Koren et~al\mbox{.}(2009)]%
        {koren2009matrix}
\bibfield{author}{\bibinfo{person}{Yehuda Koren}, \bibinfo{person}{Robert
  Bell}, {and} \bibinfo{person}{Chris Volinsky}.}
  \bibinfo{year}{2009}\natexlab{}.
\newblock \showarticletitle{Matrix factorization techniques for recommender
  systems}.
\newblock \bibinfo{journal}{\emph{Computer}} (\bibinfo{year}{2009}).
\newblock


\bibitem[Li and Tuzhilin(2020)]%
        {li2020ddtcdr}
\bibfield{author}{\bibinfo{person}{Pan Li} {and} \bibinfo{person}{Alexander
  Tuzhilin}.} \bibinfo{year}{2020}\natexlab{}.
\newblock \showarticletitle{DDTCDR: Deep dual transfer cross domain
  recommendation}. In \bibinfo{booktitle}{\emph{Proceedings of WSDM}}.
\newblock


\bibitem[Liu et~al\mbox{.}(2020)]%
        {liu2020graph}
\bibfield{author}{\bibinfo{person}{Qi Liu}, \bibinfo{person}{Ruobing Xie},
  \bibinfo{person}{Lei Chen}, \bibinfo{person}{Shukai Liu}, \bibinfo{person}{Ke
  Tu}, \bibinfo{person}{Peng Cui}, \bibinfo{person}{Bo Zhang}, {and}
  \bibinfo{person}{Leyu Lin}.} \bibinfo{year}{2020}\natexlab{}.
\newblock \showarticletitle{Graph Neural Network for Tag Ranking in
  Tag-enhanced Video Recommendation}. In \bibinfo{booktitle}{\emph{Proceedings
  of CIKM}}.
\newblock


\bibitem[Lv et~al\mbox{.}(2019)]%
        {lv2019sdm}
\bibfield{author}{\bibinfo{person}{Fuyu Lv}, \bibinfo{person}{Taiwei Jin},
  \bibinfo{person}{Changlong Yu}, \bibinfo{person}{Fei Sun},
  \bibinfo{person}{Quan Lin}, \bibinfo{person}{Keping Yang}, {and}
  \bibinfo{person}{Wilfred Ng}.} \bibinfo{year}{2019}\natexlab{}.
\newblock \showarticletitle{SDM: Sequential deep matching model for online
  large-scale recommender system}. In \bibinfo{booktitle}{\emph{Proceedings of
  CIKM}}.
\newblock


\bibitem[Man et~al\mbox{.}(2017)]%
        {man2017cross}
\bibfield{author}{\bibinfo{person}{Tong Man}, \bibinfo{person}{Huawei Shen},
  \bibinfo{person}{Xiaolong Jin}, {and} \bibinfo{person}{Xueqi Cheng}.}
  \bibinfo{year}{2017}\natexlab{}.
\newblock \showarticletitle{Cross-Domain Recommendation: An Embedding and
  Mapping Approach.}. In \bibinfo{booktitle}{\emph{Proceedings of IJCAI}}.
\newblock


\bibitem[Oord et~al\mbox{.}(2018)]%
        {oord2018representation}
\bibfield{author}{\bibinfo{person}{Aaron van~den Oord}, \bibinfo{person}{Yazhe
  Li}, {and} \bibinfo{person}{Oriol Vinyals}.} \bibinfo{year}{2018}\natexlab{}.
\newblock \showarticletitle{Representation learning with contrastive predictive
  coding}.
\newblock \bibinfo{journal}{\emph{arXiv preprint arXiv:1807.03748}}
  (\bibinfo{year}{2018}).
\newblock


\bibitem[Ouyang et~al\mbox{.}(2020)]%
        {ouyang2020minet}
\bibfield{author}{\bibinfo{person}{Wentao Ouyang}, \bibinfo{person}{Xiuwu
  Zhang}, \bibinfo{person}{Lei Zhao}, \bibinfo{person}{Jinmei Luo},
  \bibinfo{person}{Yu Zhang}, \bibinfo{person}{Heng Zou},
  \bibinfo{person}{Zhaojie Liu}, {and} \bibinfo{person}{Yanlong Du}.}
  \bibinfo{year}{2020}\natexlab{}.
\newblock \showarticletitle{MiNet: Mixed Interest Network for Cross-Domain
  Click-Through Rate Prediction}. In \bibinfo{booktitle}{\emph{Proceedings of
  CIKM}}.
\newblock


\bibitem[Paterek(2007)]%
        {paterek2007improving}
\bibfield{author}{\bibinfo{person}{Arkadiusz Paterek}.}
  \bibinfo{year}{2007}\natexlab{}.
\newblock \showarticletitle{Improving regularized singular value decomposition
  for collaborative filtering}. In \bibinfo{booktitle}{\emph{Proceedings of KDD
  cup and workshop}}.
\newblock


\bibitem[Qian et~al\mbox{.}(2020)]%
        {qian2020attribute}
\bibfield{author}{\bibinfo{person}{Tieyun Qian}, \bibinfo{person}{Yile Liang},
  \bibinfo{person}{Qing Li}, {and} \bibinfo{person}{Hui Xiong}.}
  \bibinfo{year}{2020}\natexlab{}.
\newblock \showarticletitle{Attribute Graph Neural Networks for Strict Cold
  Start Recommendation}.
\newblock \bibinfo{journal}{\emph{TKDE}} (\bibinfo{year}{2020}).
\newblock


\bibitem[Qiu et~al\mbox{.}(2020)]%
        {qiu2020gcc}
\bibfield{author}{\bibinfo{person}{Jiezhong Qiu}, \bibinfo{person}{Qibin Chen},
  \bibinfo{person}{Yuxiao Dong}, \bibinfo{person}{Jing Zhang},
  \bibinfo{person}{Hongxia Yang}, \bibinfo{person}{Ming Ding},
  \bibinfo{person}{Kuansan Wang}, {and} \bibinfo{person}{Jie Tang}.}
  \bibinfo{year}{2020}\natexlab{}.
\newblock \showarticletitle{Gcc: Graph contrastive coding for graph neural
  network pre-training}. In \bibinfo{booktitle}{\emph{Proceedings of KDD}}.
\newblock


\bibitem[Rendle(2010)]%
        {rendle2010factorization}
\bibfield{author}{\bibinfo{person}{Steffen Rendle}.}
  \bibinfo{year}{2010}\natexlab{}.
\newblock \showarticletitle{Factorization machines}. In
  \bibinfo{booktitle}{\emph{Proceedings of ICDM}}.
\newblock


\bibitem[Romero et~al\mbox{.}(2015)]%
        {romero2015fitnets}
\bibfield{author}{\bibinfo{person}{Adriana Romero}, \bibinfo{person}{Nicolas
  Ballas}, \bibinfo{person}{Samira~Ebrahimi Kahou}, \bibinfo{person}{Antoine
  Chassang}, \bibinfo{person}{Carlo Gatta}, {and} \bibinfo{person}{Yoshua
  Bengio}.} \bibinfo{year}{2015}\natexlab{}.
\newblock \showarticletitle{Fitnets: Hints for thin deep nets}. In
  \bibinfo{booktitle}{\emph{Proceedings of ICLR}}.
\newblock


\bibitem[Song et~al\mbox{.}(2019)]%
        {song2019autoint}
\bibfield{author}{\bibinfo{person}{Weiping Song}, \bibinfo{person}{Chence Shi},
  \bibinfo{person}{Zhiping Xiao}, \bibinfo{person}{Zhijian Duan},
  \bibinfo{person}{Yewen Xu}, \bibinfo{person}{Ming Zhang}, {and}
  \bibinfo{person}{Jian Tang}.} \bibinfo{year}{2019}\natexlab{}.
\newblock \showarticletitle{Autoint: Automatic feature interaction learning via
  self-attentive neural networks}. In \bibinfo{booktitle}{\emph{Proceedings of
  CIKM}}.
\newblock


\bibitem[Sun et~al\mbox{.}(2019)]%
        {sun2019bert4rec}
\bibfield{author}{\bibinfo{person}{Fei Sun}, \bibinfo{person}{Jun Liu},
  \bibinfo{person}{Jian Wu}, \bibinfo{person}{Changhua Pei},
  \bibinfo{person}{Xiao Lin}, \bibinfo{person}{Wenwu Ou}, {and}
  \bibinfo{person}{Peng Jiang}.} \bibinfo{year}{2019}\natexlab{}.
\newblock \showarticletitle{BERT4Rec: Sequential Recommendation with
  Bidirectional Encoder Representations from Transformer}. In
  \bibinfo{booktitle}{\emph{Proceedings of CIKM}}.
\newblock


\bibitem[Veli{\v{c}}kovi{\'c} et~al\mbox{.}(2018)]%
        {velivckovic2018graph}
\bibfield{author}{\bibinfo{person}{Petar Veli{\v{c}}kovi{\'c}},
  \bibinfo{person}{Guillem Cucurull}, \bibinfo{person}{Arantxa Casanova},
  \bibinfo{person}{Adriana Romero}, \bibinfo{person}{Pietro Lio}, {and}
  \bibinfo{person}{Yoshua Bengio}.} \bibinfo{year}{2018}\natexlab{}.
\newblock \showarticletitle{Graph attention networks}. In
  \bibinfo{booktitle}{\emph{Proceedings of ICLR}}.
\newblock


\bibitem[Wei et~al\mbox{.}(2021)]%
        {wei2021contrastive}
\bibfield{author}{\bibinfo{person}{Yinwei Wei}, \bibinfo{person}{Xiang Wang},
  \bibinfo{person}{Qi Li}, \bibinfo{person}{Liqiang Nie}, \bibinfo{person}{Yan
  Li}, \bibinfo{person}{Xuanping Li}, {and} \bibinfo{person}{Tat-Seng Chua}.}
  \bibinfo{year}{2021}\natexlab{}.
\newblock \showarticletitle{Contrastive Learning for Cold-Start
  Recommendation}. In \bibinfo{booktitle}{\emph{Proceedings of MM}}.
\newblock


\bibitem[Wu et~al\mbox{.}(2019)]%
        {wu2019noise}
\bibfield{author}{\bibinfo{person}{Ga Wu}, \bibinfo{person}{Maksims Volkovs},
  \bibinfo{person}{Chee~Loong Soon}, \bibinfo{person}{Scott Sanner}, {and}
  \bibinfo{person}{Himanshu Rai}.} \bibinfo{year}{2019}\natexlab{}.
\newblock \showarticletitle{Noise contrastive estimation for one-class
  collaborative filtering}. In \bibinfo{booktitle}{\emph{Proceedings of
  SIGIR}}.
\newblock


\bibitem[Wu et~al\mbox{.}(2021)]%
        {wu2021self}
\bibfield{author}{\bibinfo{person}{Jiancan Wu}, \bibinfo{person}{Xiang Wang},
  \bibinfo{person}{Fuli Feng}, \bibinfo{person}{Xiangnan He},
  \bibinfo{person}{Liang Chen}, \bibinfo{person}{Jianxun Lian}, {and}
  \bibinfo{person}{Xing Xie}.} \bibinfo{year}{2021}\natexlab{}.
\newblock \showarticletitle{Self-supervised graph learning for recommendation}.
  In \bibinfo{booktitle}{\emph{Proceedings of SIGIR}}.
\newblock


\bibitem[Wu et~al\mbox{.}(2022a)]%
        {wu2022multi}
\bibfield{author}{\bibinfo{person}{Yiqing Wu}, \bibinfo{person}{Ruobing Xie},
  \bibinfo{person}{Yongchun Zhu}, \bibinfo{person}{Xiang Ao},
  \bibinfo{person}{Xin Chen}, \bibinfo{person}{Xu Zhang},
  \bibinfo{person}{Fuzhen Zhuang}, \bibinfo{person}{Leyu Lin}, {and}
  \bibinfo{person}{Qing He}.} \bibinfo{year}{2022}\natexlab{a}.
\newblock \showarticletitle{Multi-view Multi-behavior Contrastive Learning in
  Recommendation}. In \bibinfo{booktitle}{\emph{Proceedings of DASFAA}}.
\newblock


\bibitem[Wu et~al\mbox{.}(2022b)]%
        {wu2022selective}
\bibfield{author}{\bibinfo{person}{Yiqing Wu}, \bibinfo{person}{Ruobing Xie},
  \bibinfo{person}{Yongchun Zhu}, \bibinfo{person}{Fuzhen Zhuang},
  \bibinfo{person}{Xiang Ao}, \bibinfo{person}{Xu Zhang}, \bibinfo{person}{Leyu
  Lin}, {and} \bibinfo{person}{Qing He}.} \bibinfo{year}{2022}\natexlab{b}.
\newblock \showarticletitle{Selective Fairness in Recommendation via Prompts}.
  In \bibinfo{booktitle}{\emph{Proceedings of SIGIR}}.
\newblock


\bibitem[Wu et~al\mbox{.}(2022c)]%
        {wu2022personalized}
\bibfield{author}{\bibinfo{person}{Yiqing Wu}, \bibinfo{person}{Ruobing Xie},
  \bibinfo{person}{Yongchun Zhu}, \bibinfo{person}{Fuzhen Zhuang},
  \bibinfo{person}{Xu Zhang}, \bibinfo{person}{Leyu Lin}, {and}
  \bibinfo{person}{Qing He}.} \bibinfo{year}{2022}\natexlab{c}.
\newblock \showarticletitle{Personalized Prompts for Sequential
  Recommendation}.
\newblock \bibinfo{journal}{\emph{arXiv preprint arXiv:2205.09666}}
  (\bibinfo{year}{2022}).
\newblock


\bibitem[Xiao et~al\mbox{.}(2021)]%
        {xiao2021uprec}
\bibfield{author}{\bibinfo{person}{Chaojun Xiao}, \bibinfo{person}{Ruobing
  Xie}, \bibinfo{person}{Yuan Yao}, \bibinfo{person}{Zhiyuan Liu},
  \bibinfo{person}{Maosong Sun}, \bibinfo{person}{Xu Zhang}, {and}
  \bibinfo{person}{Leyu Lin}.} \bibinfo{year}{2021}\natexlab{}.
\newblock \showarticletitle{UPRec: User-Aware Pre-training for Recommender
  Systems}.
\newblock \bibinfo{journal}{\emph{arXiv preprint arXiv:2102.10989}}
  (\bibinfo{year}{2021}).
\newblock


\bibitem[Xie et~al\mbox{.}(2021a)]%
        {xie2021improving}
\bibfield{author}{\bibinfo{person}{Ruobing Xie}, \bibinfo{person}{Qi Liu},
  \bibinfo{person}{Shukai Liu}, \bibinfo{person}{Ziwei Zhang},
  \bibinfo{person}{Peng Cui}, \bibinfo{person}{Bo Zhang}, {and}
  \bibinfo{person}{Leyu Lin}.} \bibinfo{year}{2021}\natexlab{a}.
\newblock \showarticletitle{Improving Accuracy and Diversity in Matching of
  Recommendation with Diversified Preference Network}.
\newblock \bibinfo{journal}{\emph{IEEE Transactions on Big Data}}
  (\bibinfo{year}{2021}).
\newblock


\bibitem[Xie et~al\mbox{.}(2020)]%
        {xie2020internal}
\bibfield{author}{\bibinfo{person}{Ruobing Xie}, \bibinfo{person}{Zhijie Qiu},
  \bibinfo{person}{Jun Rao}, \bibinfo{person}{Yi Liu}, \bibinfo{person}{Bo
  Zhang}, {and} \bibinfo{person}{Leyu Lin}.} \bibinfo{year}{2020}\natexlab{}.
\newblock \showarticletitle{Internal and Contextual Attention Network for
  Cold-start Multi-channel Matching in Recommendation}. In
  \bibinfo{booktitle}{\emph{Proceedings of IJCAI}}.
\newblock


\bibitem[Xie et~al\mbox{.}(2021b)]%
        {xie2021adversarial}
\bibfield{author}{\bibinfo{person}{Zhe Xie}, \bibinfo{person}{Chengxuan Liu},
  \bibinfo{person}{Yichi Zhang}, \bibinfo{person}{Hongtao Lu},
  \bibinfo{person}{Dong Wang}, {and} \bibinfo{person}{Yue Ding}.}
  \bibinfo{year}{2021}\natexlab{b}.
\newblock \showarticletitle{Adversarial and Contrastive Variational Autoencoder
  for Sequential Recommendation}. In \bibinfo{booktitle}{\emph{Proceedings of
  WWW}}.
\newblock


\bibitem[Xu et~al\mbox{.}(2018)]%
        {xu2018deep}
\bibfield{author}{\bibinfo{person}{Jun Xu}, \bibinfo{person}{Xiangnan He},
  {and} \bibinfo{person}{Hang Li}.} \bibinfo{year}{2018}\natexlab{}.
\newblock \showarticletitle{Deep learning for matching in search and
  recommendation}. In \bibinfo{booktitle}{\emph{Proceedings of SIGIR}}.
\newblock


\bibitem[You et~al\mbox{.}(2020)]%
        {you2020graph}
\bibfield{author}{\bibinfo{person}{Yuning You}, \bibinfo{person}{Tianlong
  Chen}, \bibinfo{person}{Yongduo Sui}, \bibinfo{person}{Ting Chen},
  \bibinfo{person}{Zhangyang Wang}, {and} \bibinfo{person}{Yang Shen}.}
  \bibinfo{year}{2020}\natexlab{}.
\newblock \showarticletitle{Graph contrastive learning with augmentations}.
\newblock \bibinfo{journal}{\emph{Proceedings of NeurIPS}}.
\newblock


\bibitem[Yu et~al\mbox{.}(2021)]%
        {yu2021self}
\bibfield{author}{\bibinfo{person}{Junliang Yu}, \bibinfo{person}{Hongzhi Yin},
  \bibinfo{person}{Jundong Li}, \bibinfo{person}{Qinyong Wang},
  \bibinfo{person}{Nguyen Quoc~Viet Hung}, {and} \bibinfo{person}{Xiangliang
  Zhang}.} \bibinfo{year}{2021}\natexlab{}.
\newblock \showarticletitle{Self-Supervised Multi-Channel Hypergraph
  Convolutional Network for Social Recommendation}. In
  \bibinfo{booktitle}{\emph{Proceedings of WWW}}.
\newblock


\bibitem[Zeng et~al\mbox{.}(2021)]%
        {zeng2021knowledge}
\bibfield{author}{\bibinfo{person}{Zheni Zeng}, \bibinfo{person}{Chaojun Xiao},
  \bibinfo{person}{Yuan Yao}, \bibinfo{person}{Ruobing Xie},
  \bibinfo{person}{Zhiyuan Liu}, \bibinfo{person}{Fen Lin},
  \bibinfo{person}{Leyu Lin}, {and} \bibinfo{person}{Maosong Sun}.}
  \bibinfo{year}{2021}\natexlab{}.
\newblock \showarticletitle{Knowledge transfer via pre-training for
  recommendation: A review and prospect}.
\newblock \bibinfo{journal}{\emph{Frontiers in big Data}}
  (\bibinfo{year}{2021}).
\newblock


\bibitem[Zhao et~al\mbox{.}(2019)]%
        {zhao2019cross}
\bibfield{author}{\bibinfo{person}{Cheng Zhao}, \bibinfo{person}{Chenliang Li},
  {and} \bibinfo{person}{Cong Fu}.} \bibinfo{year}{2019}\natexlab{}.
\newblock \showarticletitle{Cross-domain recommendation via preference
  propagation GraphNet}. In \bibinfo{booktitle}{\emph{Proceedings of CIKM}}.
\newblock


\bibitem[Zhao et~al\mbox{.}(2020)]%
        {zhao2020catn}
\bibfield{author}{\bibinfo{person}{Cheng Zhao}, \bibinfo{person}{Chenliang Li},
  \bibinfo{person}{Rong Xiao}, \bibinfo{person}{Hongbo Deng}, {and}
  \bibinfo{person}{Aixin Sun}.} \bibinfo{year}{2020}\natexlab{}.
\newblock \showarticletitle{CATN: Cross-domain recommendation for cold-start
  users via aspect transfer network}. In \bibinfo{booktitle}{\emph{Proceedings
  of SIGIR}}.
\newblock


\bibitem[Zhou et~al\mbox{.}(2021)]%
        {zhou2021contrastive}
\bibfield{author}{\bibinfo{person}{Chang Zhou}, \bibinfo{person}{Jianxin Ma},
  \bibinfo{person}{Jianwei Zhang}, \bibinfo{person}{Jingren Zhou}, {and}
  \bibinfo{person}{Hongxia Yang}.} \bibinfo{year}{2021}\natexlab{}.
\newblock \showarticletitle{Contrastive Learning for Debiased Candidate
  Generation in Large-Scale Recommender Systems}. In
  \bibinfo{booktitle}{\emph{Proceedings of KDD}}.
\newblock


\bibitem[Zhou et~al\mbox{.}(2020)]%
        {zhou2020s3}
\bibfield{author}{\bibinfo{person}{Kun Zhou}, \bibinfo{person}{Hui Wang},
  \bibinfo{person}{Wayne~Xin Zhao}, \bibinfo{person}{Yutao Zhu},
  \bibinfo{person}{Sirui Wang}, \bibinfo{person}{Fuzheng Zhang},
  \bibinfo{person}{Zhongyuan Wang}, {and} \bibinfo{person}{Ji-Rong Wen}.}
  \bibinfo{year}{2020}\natexlab{}.
\newblock \showarticletitle{S3-rec: Self-supervised learning for sequential
  recommendation with mutual information maximization}. In
  \bibinfo{booktitle}{\emph{Proceedings of CIKM}}.
\newblock


\bibitem[Zhu et~al\mbox{.}(2018)]%
        {zhu2018learning}
\bibfield{author}{\bibinfo{person}{Han Zhu}, \bibinfo{person}{Xiang Li},
  \bibinfo{person}{Pengye Zhang}, \bibinfo{person}{Guozheng Li},
  \bibinfo{person}{Jie He}, \bibinfo{person}{Han Li}, {and}
  \bibinfo{person}{Kun Gai}.} \bibinfo{year}{2018}\natexlab{}.
\newblock \showarticletitle{Learning Tree-based Deep Model for Recommender
  Systems}. In \bibinfo{booktitle}{\emph{Proceedings of KDD}}.
\newblock


\bibitem[Zhu et~al\mbox{.}(2021)]%
        {zhu2021transfer}
\bibfield{author}{\bibinfo{person}{Yongchun Zhu}, \bibinfo{person}{Kaikai Ge},
  \bibinfo{person}{Fuzhen Zhuang}, \bibinfo{person}{Ruobing Xie},
  \bibinfo{person}{Dongbo Xi}, \bibinfo{person}{Xu Zhang},
  \bibinfo{person}{Leyu Lin}, {and} \bibinfo{person}{Qing He}.}
  \bibinfo{year}{2021}\natexlab{}.
\newblock \showarticletitle{Transfer-Meta Framework for Cross-domain
  Recommendation to Cold-Start Users}. In \bibinfo{booktitle}{\emph{Proceedings
  of SIGIR}}.
\newblock


\bibitem[Zhu et~al\mbox{.}(2022)]%
        {zhu2022personalized}
\bibfield{author}{\bibinfo{person}{Yongchun Zhu}, \bibinfo{person}{Zhenwei
  Tang}, \bibinfo{person}{Yudan Liu}, \bibinfo{person}{Fuzhen Zhuang},
  \bibinfo{person}{Ruobing Xie}, \bibinfo{person}{Xu Zhang},
  \bibinfo{person}{Leyu Lin}, {and} \bibinfo{person}{Qing He}.}
  \bibinfo{year}{2022}\natexlab{}.
\newblock \showarticletitle{Personalized transfer of user preferences for
  cross-domain recommendation}. In \bibinfo{booktitle}{\emph{Proceedings of
  WSDM}}.
\newblock


\end{thebibliography}

\appendix

\section{Appendix}

\subsection{Online Deployment and Efficiency}
\label{app:online_deployment}

\subsubsection{Online Deployment}

We have deployed CCDR on WeChat Top Stories (a well-known recommendation system widely used by millions of users) for more than six months.
In practical systems, the quality, novelty, and diversity of item contents are the dominating factors that strongly impact user experiences. Hence, real-world recommendation systems often need to introduce new data sources as cold-start domains to add more contents and attract users.
CCDR is deployed on the matching module to rapidly retrieve high-quality items. The relatively few-shot domain is viewed as the target domain, while other many-shot domains are combined as the source domain. We retrieve top $500$ items for each target domain in matching. These retrieved items of CCDR are combined with candidates of other domains, and are then fed into the following ranking module for the final display. The computation cost is acceptable for the offline daily training in practical systems.
CCDR is a straightforward, effective, and efficient implementation for CDR in matching, which can also be easily adopted in other new domains.

\subsubsection{Online Efficiency}

Real-world matching should deal with more than million-level candidates, where a linear prediction complexity w.r.t. the huge corpus size is unacceptable \cite{zhu2018learning}. In CCDR, all embedding-based retrievals are conducted via the node embedding similarities learned in Eq. (\ref{eq.L_all}), which can be pre-calculated in offline as stated in Sec. \ref{sec.online_deployment}. The top-100 nearest items for all heterogeneous nodes are also pre-indexed in offline for efficiency.
Therefore, we do not need to calculate $sim(\cdot,\cdot)$ in online, and the online computation cost mainly locates in the indexing (finding top-100 nearest item embeddings via indexes) and sorting part (selecting the final top-500 items via $s_i$) of Eq. (\ref{eq.item_channel}) to Eq. (\ref{eq.all_score}). We have an industry-level online indexer with cache technologies, which could largely accelerate the indexing part to nearly $O(1)$ and make sure it won't become a computation bottleneck. Since we have six channels and select top-$100$ nearest items for each heterogeneous object, the computation complexity of the indexing part is nearly $O(600n)$ ($n$ is the length of the user historical behavior, and the maximum user behavior length $n$ is $200$). Similarly, the sorting part is $O(\log(600n))$. In conclusion, the overall online computation complexity is acceptable, and contrastive learning won't bring in additional online computation.
As for the memory cost, CCDR only needs space to store the top-N nearest neighbor indexes for all nodes, which is acceptable for an industrial large-scale recommendation system.

\subsection{Dataset}
\label{app:dataset}

We build the CDR-427M dataset from a real-world multi-domain recommendation system in WeChat Top Stories. Specifically, we select two relatively new domains as the target domains, including a few-shot target domain where each user may have several behaviors, and a strict cold-start target domain where all users have no target-domain behaviors in the train set \cite{qian2020attribute}. Users with the same account in different domains are aligned naturally.
To simulate real-world scenarios and make full use of all information, other many-shot domains existing in our system (the main feed) are combined as the source domain. Hence, most users are overlapped in the global large source domain.
Note that we evaluate on the strict cold-start setting to simulate the practical scenario when some new data sources are added to the existing recommendation system for the first time (which is not rare in the industry). The users are randomly selected and the dataset is split according to the chronological order. The item attributes used in CCDR help knowledge transfer, which can be obtained in most recommender systems. The inter-CL is universal.
All data are preprocessed via data masking to protect the user’s privacy.

We build the diversified preference networks for three domains separately following Sec. \ref{sec.heterogeneous_network}. The tags and categories are shared across different domains. Each item has one media (i.e., the item's producer). The words are extracted from the items' titles. These four types of item attributes widely exist and could be easily collected in most industrial recommendation systems. Besides user-item interactions, we further consider the session information to build item behavioral correlations. In total, there are $2.49$, $0.5$, $0.73$ million nodes and $175$, $29.8$, $15.0$ million edges in the diversified preference networks of the source domain, few-shot target domain, and strict cold-start domain, respectively.
We bring in heterogeneous information to (a) better learn node representations in each domain besides sparse user-item interactions, and (b) make full use of all possible knowledge transfer paths between different domains. In fact, CCDR has been widely used in multiple online domains.

\subsection{Motivations and Implementation Details}
\label{app:implementation}

\subsubsection{Motivations of Model Designs}

In this work, we start from the challenging cold-start issue when adding new data sources or new recommendation scenarios, which often happens in practical recommendations. Cross-domain recommendation is a promising solution to this issue. However, it is trivial and time-consuming to design effective customized domain-specific models for different target domains. Hence, our proposed CDR model should not only be \textbf{effective} and \textbf{efficient}, but also be \textbf{universal} and \textbf{easy to deploy}. We should double clarify that the \emph{online performance} is the central objective. Therefore, rather than fancy complicated methods, we prefer \textbf{simple} and effective models according to our industry taste.

Under these principles, we first bring in contrastive learning to improve representation learning and knowledge transfer in CDR with different intra-CL and inter-CL tasks, which are two of our main contributions. We bring CL into CDR of matching, since CL-based self-supervised training enables more sufficient training especially for those long-tail and unaligned items in the target domain.
We do not propose new CL-based architectures or design fancy complicated GNN aggregators, since we find that the classical InfoNCE and GAT-based aggregation have already achieved significant offline and online improvements in different domains. Instead, we propose new types of augmentations for CDR, which are also an essential component in CL.
For GNN aggregator, we directly use GAT, since we suggest that all heterogeneous embeddings are in the same semantic space, as we should retrieve diversified items via similarities between different heterogeneous objects in online multi-channel matching. The single-domain GNN modeling is not the main focus of this work. Nevertheless, it is still convenient to conduct other heterogeneous GNN models in node aggregation.
Although more sophisticated CL tasks and GNN models may further improve the performance, the universality, offline/online efficiency, and deployment costs will be worse. How to balance the four real-world principles with enhanced models could be our future work.
For the construction of the diversified preference network, we follow the existing model's setting \cite{xie2021improving} and modify it according to our system's characteristics. It is also fine to use other informative interactions for more sufficient training.

As for the specific advantages and motivations of different components, the advantages of CCDR are introduced in Sec. \ref{sec.introduction} and Sec. \ref{sec.offline_evaluation} with evaluation results. The advantages of the diversified preference network and the information brought by different edges are given in Sec. \ref{sec.heterogeneous_network}. The motivations of both intra-CL and inter-CL are in Sec. \ref{sec.intra_CL}, Sec. \ref{sec.inter_CL}, and Sec. \ref{sec.ablation}. The feasibility and advantages of the cross-domain multi-channel matching are also concluded in Sec. \ref{sec.online_deployment}. Different components and model designs cooperate for the same goals of improving representation learning and cross-domain knowledge transfer, aiming to alleviate the sparsity issues in CDR.

\subsubsection{Implementation Details}

Sec. \ref{sec.experimental_settings} has shown most essential hyper-parameters of CCDR. For the basic parameters such as the input and output dimensions of heterogeneous node embeddings and the numbers of weighted neighbor sampling, we empirically set them according to our existing online GNN-based matching models to balance effectiveness and efficiency under the practical computation limitation.
We conduct a grid search to select sensitive parameters. The most essential parameters are the weights of different losses. For the CL loss weights, we have chosen different intra-CL and inter-CL loss weights among $\{0.2, 0.6, 1.0, 1.5, 2.0\}$, and find the best parameters as $\lambda_1,\lambda_2,\lambda_3,\lambda_4=1.0,1.0,1.5,0.6$. Fig. \ref{fig:paramater} shows the results of different loss weights. For the possible channel weights in online cross-domain multi-channel matching, we empirically use the online default setting (i.e., equal channel weights). It is also easy to adjust the heterogeneous channels’ weights for practical demands (e.g., if we want to highlight user's media-level preferences, we can choose bigger media channel weights), which is not the main concern of this work. For the batch size, we have tested among $\{128, 512, 1024, 4096\}$, and select the best $4,096$ with larger batch size \cite{chen2020simple}.

\subsection{Evaluation Details}
\label{app:evaluation}

\subsubsection{Baselines}

We focus on cross-domain recommendation in matching. Therefore, we conduct three groups of competitive baselines for comprehensive comparisons.
The first group is the classical matching methods, including three classical industrial matching models (i.e., FM, AutoInt, and GraphDR+). Note that both the source and target domains' behaviors are considered in these baselines for fair comparisons. GraphDR \cite{xie2021improving} is the SOTA graph-based matching model. We enhance the original GraphDR model for the CDR task as GraphDR+, which also acts as the single-domain model of our CCDR for fair comparisons. We do not compare with other matching models like ComiRec \cite{cen2020controllable} and TDM \cite{zhu2018learning}, for they are designed for different challenges and are nontrivial to be modified for CDR.
The second group is cross-domain/multi-domain methods, including EMCDR, SSCDR, and ICAN. For EMCDR and SSCDR, we also adopt GraphDR+ as the single-domain model. ICAN is the SOTA multi-domain matching model. We also do not compare with CoNet \cite{hu2018conet} and CATN \cite{zhao2020catn}, for they are designed for ranking and will be extremely time-consuming in matching.
The third group contains another knowledge transfer method of knowledge distillation, which is widely used in practice. Moreover, existing graph-based intra-CL techniques \cite{qiu2020gcc,wu2021self} are also implemented as baselines and ablation versions.

\subsubsection{Offline Evaluation Metrics}

We concentrate on the CDR in matching. In practical large-scale recommendation systems, there are usually million-level item candidates, where complicated user-item feature interaction modeling is hard to be calculated on the overall corpus. Hence, the two-step recommendation system is proposed \cite{covington2016deep}, where the matching module focuses more on efficiently retrieving a small subset of item candidates from the million-level large corpora, while the ranking module aims to give the specific top-k item ranks accurately.
In our online system, we select top-500 items in the matching module for the following ranking. The specific item ranks of these top-500 items are not important, since they will be re-ranked in the following ranking and mixing modules after matching. Hence, following classical matching models \cite{sun2019bert4rec,xie2020internal,xie2021improving}, we use the top $N$ hit rate (HIT@N) as our evaluation metric instead of ranking metrics such as NDCG and AUC. To simulate the real-world matching systems, we report the HIT@N with larger $N$ as $50$, $100$, $200$, and $500$.
Moreover, since matching often struggles with the popularity issue, we also evaluate the recommendation diversity via a classical aggregate diversity metric named item coverage \cite{herlocker2004evaluating}. The item coverage measures the percentage of a dataset that
the recommendation model is able to provide predictions for in the test set.
In conclusion, we find significant improvements in both recommendation accuracy and diversity.
Note that although we do not explicitly optimize diversity, CCDR still has $4.2\%$ and $6.0\%$ improvements on item coverage over GraphDR+.

\subsubsection{Online Evaluation Metrics}

In online evaluation, we focus on three representative online metrics in the target domain: (1) Click-through-rate (CTR), (2) average user duration per capita (Duration), and (3) average share rate per capita (Share rate). We have:
\begin{equation}
\begin{split}
&\mathrm{CTR}=\frac{\mathrm{\#\ of\ click\ counts}}{\mathrm{\#\ of\ impression\ counts}},\quad \quad
\mathrm{Duration}=\frac{\mathrm{all\ duration}}{\mathrm{\#\ of\ users}},\\
&\mathrm{Share\ rate}=\frac{\mathrm{\#\ of\ share\ counts}}{\mathrm{\#\ of\ users}}.
\end{split}
\end{equation}
CTR measures the click-related recommendation accuracy. Duration focuses on the dwell time, which can better reflect user's real satisfaction. The share rate can reflect user's satisfaction via a more high-cost user social behavior. We conduct the A/B test for $8$ days.

\end{document}